\documentclass[12pt]{article}
\usepackage{epsfig}
\usepackage{feynarts}
\usepackage{axodraw}
\newcommand{\agt}{\,\rlap{\lower 3.5 pt \hbox{$\mathchar \sim$}} \raise 1pt
 \hbox {$>$}\,}
\newcommand{\alt}{\,\rlap{\lower 3.5 pt \hbox{$\mathchar \sim$}} \raise 1pt
 \hbox {$<$}\,}

\catcode`@=11
\newcount\@tempcntc
\def\@citex[#1]#2{\if@filesw\immediate\write\@auxout{\string\citation{#2}}\fi
  \@tempcnta\z@\@tempcntb\m@ne\def\@citea{}\@cite{\@for\@citeb:=#2\do
    {\@ifundefined
       {b@\@citeb}{\@citeo\@tempcntb\m@ne\@citea\def\@citea{,}{\bf ?}\@warning
       {Citation `\@citeb' on page \thepage \space undefined}}%
    {\setbox\z@\hbox{\global\@tempcntc0\csname b@\@citeb\endcsname\relax}%
     \ifnum\@tempcntc=\z@ \@citeo\@tempcntb\m@ne
       \@citea\def\@citea{,}\hbox{\csname b@\@citeb\endcsname}%
     \else
      \advance\@tempcntb\@ne
      \ifnum\@tempcntb=\@tempcntc
      \else\advance\@tempcntb\m@ne\@citeo
      \@tempcnta\@tempcntc\@tempcntb\@tempcntc\fi\fi}}\@citeo}{#1}}
\def\@citeo{\ifnum\@tempcnta>\@tempcntb\else\@citea\def\@citea{,}%
  \ifnum\@tempcnta=\@tempcntb\the\@tempcnta\else
   {\advance\@tempcnta\@ne\ifnum\@tempcnta=\@tempcntb \else \def\@citea{--}\fi
    \advance\@tempcnta\m@ne\the\@tempcnta\@citea\the\@tempcntb}\fi\fi}
\catcode`@=12

\begin{document}

\title{
\vskip-3cm{\baselineskip14pt
\centerline{\normalsize DESY 06-041\hfill ISSN 0418-9833}
\centerline{\normalsize MPP-2006-57\hfill}
\centerline{\normalsize hep-ph/0606073\hfill}
\centerline{\normalsize June 2006\hfill}
}
\vskip1.5cm
Single top production at HERA in the Standard Model and its minimal
supersymmetric extension}
\author{
{\sc W. Hollik${}^a$, A. H\"uttmann${}^b$\thanks{Present address: Deutsches
Elektronen-Synchrotron (DESY), Notkestr.\ 85, 22607 Hamburg, Germany.},
B.A. Kniehl${}^b$}\\
{\normalsize ${}^a$ Max-Planck-Institut f\"ur Physik
(Werner-Heisenberg-Institut),}\\
{\normalsize F\"ohringer Ring 6, 80805 Munich, Germany}\\
{\normalsize ${}^b$ II. Institut f\"ur Theoretische Physik,
Universit\"at Hamburg,}\\
{\normalsize Luruper Chaussee 149, 22761 Hamburg, Germany}}

\date{Reveived 7 June 2006}

\maketitle

\thispagestyle{empty}

\begin{abstract}
The H1 Collaboration at the DESY electron-proton collider HERA has observed,
in photoproduction and neutral-current deep-inelastic scattering, an
unexpected excess of events with isolated leptons and missing transverse
momentum, especially at large values of hadronic transverse momentum---a
signature typical for single top-quark production.
This observation is being substantiated in the HERA~II run.
Motivated by this, we evaluate the cross section of single top-quark photo-
and electroproduction both in the standard model and its minimal
supersymmetric extension, considering both minimal and non-minimal
flavour-violation scenarios in the latter case.

\medskip

\end{abstract}

\newpage

\section{Introduction}
\label{sec:one}

Searches for single top-quark production via the neutral current (NC),
\begin{equation}
e^\pm p\rightarrow e^\pm t+X,
\label{eq:had}
\end{equation}
have been performed by the H1 Collaboration \cite{H1} and the ZEUS
Collaboration \cite{ZEUS} at the DESY electron-proton collider HERA.
The H1 Collaboration found several events, leading to a cross section of
$\sigma= 0.29{+0.15\atop-0.14}$~pb.
Alternatively, assuming that the observed events are due to a statistical
fluctuation, upper limits of 0.55~pb on $\sigma$ and of 0.27 on the anomalous
$tu\gamma$ coupling $\kappa_{tu\gamma}$ were established at the 95\% confidence
level (CL).
On the other hand, the ZEUS Collaboration found no evidence for top-quark
production and was able to place upper bounds of 0.225~pb on $\sigma$ and of
0.174 on $\kappa_{tu\gamma}$ at 95\% CL.

Single top-quark production via the charged current (CC) is possible at the
tree level, but its cross section is less than 1~fb \cite{CC}.
Furthermore, such events can be separated experimentally due to the absence of
a scattered electron or positron in the final state.

In this work, we calculate the cross section of process~(\ref{eq:had}) in the
standard model (SM) and its minimal supersymmetric (SUSY) extension (MSSM),
considering both scenarios with minimal flavour violation (MFV) and
non-minimal flavour violation (NMFV).
The bulk of the cross section is due to photoproduction and electromagnetic
deep-inelastic scattering (DIS), while the contribution due to the exchange of
a virtual $Z$ boson is greatly suppressed by its mass.
Since photonic interactions cannot change flavour and the top quark does not
appear as a parton in the proton, we are dealing here with a loop-induced
process.
In the SM and the MFV MSSM, its cross section is further suppressed by the
smallness of the contributing elements of the Cabibbo-Kobayashi-Maskawa (CKM)
matrix.
In more general MSSM scenarios, misalignment between the quark and squark
sectors can appear, and the CKM matrix is no longer the only source of flavour
violation.
Thus, the flavour-changing (FC) couplings are not Cabibbo suppressed, and
sizeable contributions to FC NC processes can occur.
On the other hand, there are strong experimental bounds on squark mixing
involving the first generation, coming from data on $K^0$--$\overline{K}^0$
and $D^0$--$\overline{D}^0$ mixing \cite{K0D0}.
If the squark mixing involving the first generation is neglected completely,
there are no new vertices for incoming up-quarks compared to the MFV MSSM.
Thus, every new contribution in this scenario is suppressed by the parton
distribution function (PDF) of the charm quark in the proton.
Because the SM cross section for NC single top-quark production is highly
suppressed, every detected event is an indication of physics beyond the SM.

This paper is organised as follows.
In Section~\ref{sec:two}, we describe the analytical calculation of the SM
cross section.
In Section~\ref{sec:three}, we outline the theoretical framework of FC
interactions in the MSSM.
The numerical analysis is presented in Section~\ref{sec:four}.
Our conclusions are summarised in Section~\ref{sec:five}. 

\section{SM cross section}
\label{sec:two}

In this section, the analytical calculation of the SM cross section is
described.
As indicated in Fig.~\ref{overview}, we denote the four-momenta of the
incoming proton, parton (up or charm quark), and electron by $P$, $p$, and
$k$, respectively, and those of the outgoing top quark and electron by
$p^\prime$ and $k^\prime$, respectively.
As for the centre-of-mass (CM) energy and the top-quark mass, we have
$S=(P+k)^2$ and $m_t^2=(p^\prime)^2$.
We neglect the masses of the proton, the incoming quarks, and the electron, so
that $P^2=p^2=k^2=k^{\prime2}=0$.
The four-momentum of the exchanged photon is given by $q=k-k^\prime$, and, as
usual, we introduce the virtuality variable $Q^2=-q^2>0$.
The variable $y=(q\cdot P)/(k\cdot P)$ measures the relative electron energy
loss in the proton rest frame.

\subsection{Electroproduction}

\begin{center}
\begin{figure}
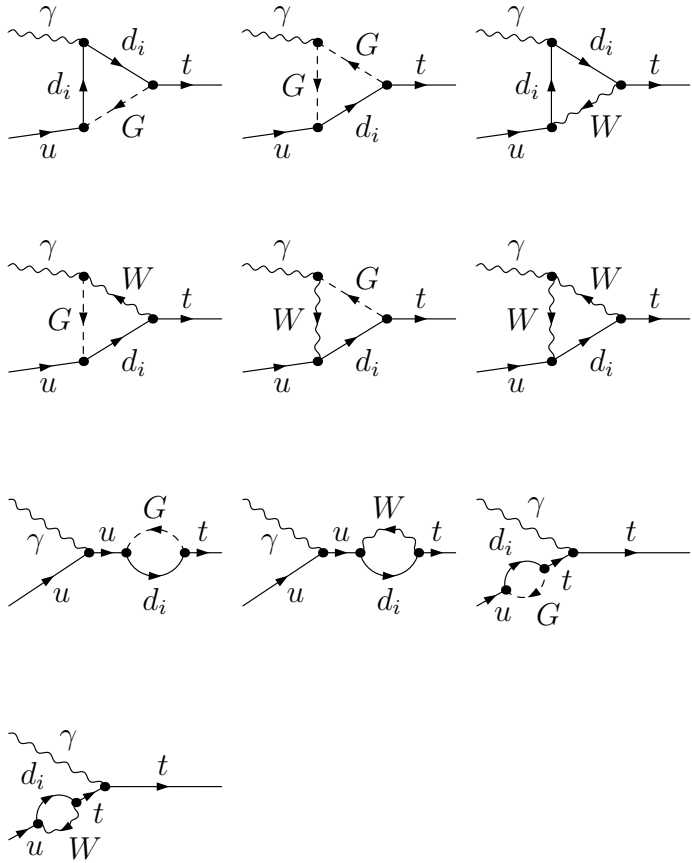


\unitlength=0.7bp%

\begin{feynartspicture}(432,504)(3,4)

\FADiagram{}
\FAProp(0.,15.)(7.,14.)(0.,){/Sine}{0}
\FALabel(3.7192,15.5544)[b]{$\gamma$}
\FAProp(0.,5.)(7.,6.)(0.,){/Straight}{1}
\FALabel(3.7192,4.44558)[t]{$u$}
\FAProp(20.,10.)(13.5,10.)(0.,){/Straight}{-1}
\FALabel(16.75,11.07)[b]{$t$}
\FAProp(7.,14.)(7.,6.)(0.,){/Straight}{-1}
\FALabel(5.93,10.)[r]{$d_i$}
\FAProp(7.,14.)(13.5,10.)(0.,){/Straight}{1}
\FALabel(10.5824,12.8401)[bl]{$d_i$}
\FAProp(7.,6.)(13.5,10.)(0.,){/ScalarDash}{-1}
\FALabel(10.5824,7.15993)[tl]{$G$}
\FAVert(7.,14.){0}
\FAVert(7.,6.){0}
\FAVert(13.5,10.){0}

\FADiagram{}
\FAProp(0.,15.)(7.,14.)(0.,){/Sine}{0}
\FALabel(3.7192,15.5544)[b]{$\gamma$}
\FAProp(0.,5.)(7.,6.)(0.,){/Straight}{1}
\FALabel(3.7192,4.44558)[t]{$u$}
\FAProp(20.,10.)(13.5,10.)(0.,){/Straight}{-1}
\FALabel(16.75,11.07)[b]{$t$}
\FAProp(7.,14.)(7.,6.)(0.,){/ScalarDash}{1}
\FALabel(5.93,10.)[r]{$G$}
\FAProp(7.,14.)(13.5,10.)(0.,){/ScalarDash}{-1}
\FALabel(10.5824,12.8401)[bl]{$G$}
\FAProp(7.,6.)(13.5,10.)(0.,){/Straight}{1}
\FALabel(10.5824,7.15993)[tl]{$d_i$}
\FAVert(7.,14.){0}
\FAVert(7.,6.){0}
\FAVert(13.5,10.){0}

\FADiagram{}
\FAProp(0.,15.)(7.,14.)(0.,){/Sine}{0}
\FALabel(3.7192,15.5544)[b]{$\gamma$}
\FAProp(0.,5.)(7.,6.)(0.,){/Straight}{1}
\FALabel(3.7192,4.44558)[t]{$u$}
\FAProp(20.,10.)(13.5,10.)(0.,){/Straight}{-1}
\FALabel(16.75,11.07)[b]{$t$}
\FAProp(7.,14.)(7.,6.)(0.,){/Straight}{-1}
\FALabel(5.93,10.)[r]{$d_i$}
\FAProp(7.,14.)(13.5,10.)(0.,){/Straight}{1}
\FALabel(10.5824,12.8401)[bl]{$d_i$}
\FAProp(7.,6.)(13.5,10.)(0.,){/Sine}{-1}
\FALabel(10.5824,7.15993)[tl]{$W$}
\FAVert(7.,14.){0}
\FAVert(7.,6.){0}
\FAVert(13.5,10.){0}

\FADiagram{}
\FAProp(0.,15.)(7.,14.)(0.,){/Sine}{0}
\FALabel(3.7192,15.5544)[b]{$\gamma$}
\FAProp(0.,5.)(7.,6.)(0.,){/Straight}{1}
\FALabel(3.7192,4.44558)[t]{$u$}
\FAProp(20.,10.)(13.5,10.)(0.,){/Straight}{-1}
\FALabel(16.75,11.07)[b]{$t$}
\FAProp(7.,14.)(7.,6.)(0.,){/ScalarDash}{1}
\FALabel(5.93,10.)[r]{$G$}
\FAProp(7.,14.)(13.5,10.)(0.,){/Sine}{-1}
\FALabel(10.5824,12.8401)[bl]{$W$}
\FAProp(7.,6.)(13.5,10.)(0.,){/Straight}{1}
\FALabel(10.5824,7.15993)[tl]{$d_i$}
\FAVert(7.,14.){0}
\FAVert(7.,6.){0}
\FAVert(13.5,10.){0}

\FADiagram{}
\FAProp(0.,15.)(7.,14.)(0.,){/Sine}{0}
\FALabel(3.7192,15.5544)[b]{$\gamma$}
\FAProp(0.,5.)(7.,6.)(0.,){/Straight}{1}
\FALabel(3.7192,4.44558)[t]{$u$}
\FAProp(20.,10.)(13.5,10.)(0.,){/Straight}{-1}
\FALabel(16.75,11.07)[b]{$t$}
\FAProp(7.,14.)(7.,6.)(0.,){/Sine}{1}
\FALabel(5.93,10.)[r]{$W$}
\FAProp(7.,14.)(13.5,10.)(0.,){/ScalarDash}{-1}
\FALabel(10.5824,12.8401)[bl]{$G$}
\FAProp(7.,6.)(13.5,10.)(0.,){/Straight}{1}
\FALabel(10.5824,7.15993)[tl]{$d_i$}
\FAVert(7.,14.){0}
\FAVert(7.,6.){0}
\FAVert(13.5,10.){0}

\FADiagram{}
\FAProp(0.,15.)(7.,14.)(0.,){/Sine}{0}
\FALabel(3.7192,15.5544)[b]{$\gamma$}
\FAProp(0.,5.)(7.,6.)(0.,){/Straight}{1}
\FALabel(3.7192,4.44558)[t]{$u$}
\FAProp(20.,10.)(13.5,10.)(0.,){/Straight}{-1}
\FALabel(16.75,11.07)[b]{$t$}
\FAProp(7.,14.)(7.,6.)(0.,){/Sine}{1}
\FALabel(5.93,10.)[r]{$W$}
\FAProp(7.,14.)(13.5,10.)(0.,){/Sine}{-1}
\FALabel(10.5824,12.8401)[bl]{$W$}
\FAProp(7.,6.)(13.5,10.)(0.,){/Straight}{1}
\FALabel(10.5824,7.15993)[tl]{$d_i$}
\FAVert(7.,14.){0}
\FAVert(7.,6.){0}
\FAVert(13.5,10.){0}

\FADiagram{}
\FAProp(0.,15.)(7.5,10.)(0.,){/Sine}{0}
\FALabel(3.37021,11.6903)[tr]{$\gamma$}
\FAProp(0.,5.)(7.5,10.)(0.,){/Straight}{1}
\FALabel(4.12979,6.69032)[tl]{$u$}
\FAProp(20.,10.)(16.5,10.)(0.,){/Straight}{-1}
\FALabel(18.25,11.07)[b]{$t$}
\FAProp(7.5,10.)(11.,10.)(0.,){/Straight}{1}
\FALabel(9.25,11.07)[b]{$u$}
\FAProp(16.5,10.)(11.,10.)(-0.8,){/Straight}{-1}
\FALabel(13.75,6.73)[t]{$d_i$}
\FAProp(16.5,10.)(11.,10.)(0.8,){/ScalarDash}{1}
\FALabel(13.75,13.27)[b]{$G$}
\FAVert(7.5,10.){0}
\FAVert(16.5,10.){0}
\FAVert(11.,10.){0}

\FADiagram{}
\FAProp(0.,15.)(7.5,10.)(0.,){/Sine}{0}
\FALabel(3.37021,11.6903)[tr]{$\gamma$}
\FAProp(0.,5.)(7.5,10.)(0.,){/Straight}{1}
\FALabel(4.12979,6.69032)[tl]{$u$}
\FAProp(20.,10.)(16.5,10.)(0.,){/Straight}{-1}
\FALabel(18.25,11.07)[b]{$t$}
\FAProp(7.5,10.)(11.,10.)(0.,){/Straight}{1}
\FALabel(9.25,11.07)[b]{$u$}
\FAProp(16.5,10.)(11.,10.)(-0.8,){/Straight}{-1}
\FALabel(13.75,6.73)[t]{$d_i$}
\FAProp(16.5,10.)(11.,10.)(0.8,){/Sine}{1}
\FALabel(13.75,13.27)[b]{$W$}
\FAVert(7.5,10.){0}
\FAVert(16.5,10.){0}
\FAVert(11.,10.){0}

\FADiagram{}
\FAProp(0.,15.)(9.,10.)(0.,){/Sine}{0}
\FALabel(4.77275,13.3749)[bl]{$\gamma$}
\FAProp(0.,5.)(2.7,6.5)(0.,){/Straight}{1}
\FALabel(1.62275,4.87506)[tl]{$u$}
\FAProp(20.,10.)(9.,10.)(0.,){/Straight}{-1}
\FALabel(14.5,11.07)[b]{$t$}
\FAProp(9.,10.)(6.3,8.5)(0.,){/Straight}{-1}
\FALabel(7.92275,8.37506)[tl]{$t$}
\FAProp(2.7,6.5)(6.3,8.5)(-0.8,){/Straight}{1}
\FALabel(3.42725,9.81494)[br]{$d_i$}
\FAProp(2.7,6.5)(6.3,8.5)(0.8,){/ScalarDash}{-1}
\FALabel(5.57275,5.18506)[tl]{$G$}
\FAVert(9.,10.){0}
\FAVert(2.7,6.5){0}
\FAVert(6.3,8.5){0}

\FADiagram{}
\FAProp(0.,15.)(9.,10.)(0.,){/Sine}{0}
\FALabel(4.77275,13.3749)[bl]{$\gamma$}
\FAProp(0.,5.)(2.7,6.5)(0.,){/Straight}{1}
\FALabel(1.62275,4.87506)[tl]{$u$}
\FAProp(20.,10.)(9.,10.)(0.,){/Straight}{-1}
\FALabel(14.5,11.07)[b]{$t$}
\FAProp(9.,10.)(6.3,8.5)(0.,){/Straight}{-1}
\FALabel(7.92275,8.37506)[tl]{$t$}
\FAProp(2.7,6.5)(6.3,8.5)(-0.8,){/Straight}{1}
\FALabel(3.42725,9.81494)[br]{$d_i$}
\FAProp(2.7,6.5)(6.3,8.5)(0.8,){/Sine}{-1}
\FALabel(5.57275,5.18506)[tl]{$W$}
\FAVert(9.,10.){0}
\FAVert(2.7,6.5){0}
\FAVert(6.3,8.5){0}








\end{feynartspicture}

\caption{\label{FeynmanSM}Feynman diagrams of partonic
subprocess~(\ref{eq:sub}) in the SM.
Here, $G$ is the charged Goldstone boson and $d_i$ with $i=1,2,3$ are the
down-type quarks.}

\end{figure}
\end{center}

Single top-quark production in NC DIS occurs via the partonic subprocess
\begin{equation}
e^\pm q \rightarrow e^\pm t,
\label{eq:sub}
\end{equation}
where $q=u,c$.
The Feynman diagrams contributing in the SM to process~(\ref{eq:sub}) with
$q=u$ are depicted in Fig.~\ref{FeynmanSM}.
The ones for $q=c$ are similar.
The amplitude of this processes was also calculated in Ref.~\cite{SMMatr}.
There appears at least one off-diagonal element of the CKM matrix in each
term.
If the contributions of the inner-quark flavours to a single Feynman diagram
are added up and their masses are neglected, the amplitude of this Feynman
diagram vanishes due to the unitarity of the CKM matrix.
Thus, we cannot neglect the inner-quark masses here.
Although our choice $m_c=0<m_d$ appears unphysical at first sight, it is
inconsequential in practice.
In fact, we verified that the use of a realistic value of $m_c$ affects our
numerical results only insignificantly.
A detailed proof of hard-scattering factorisation with the inclusion of
heavy-quark masses may be found in Ref.~\cite{col}.

Our calculation proceeds along the lines of Ref.~\cite{Lennart}.
The differential cross section of process~(\ref{eq:sub}) reads
\begin{equation}
\left(\frac{d\sigma}{dQ^2 dy}\right)_{\rm part}
=\frac{1}{16\pi \xi S}\overline{{\left|\mathcal{M}\right|}^2}
\delta\left(\xi y S-Q^2-m_t^2\right),
\label{eq:xs}
\end{equation}
where $\xi$, defined as $p=\xi P$, is the fraction of the proton momentum
passed on to the incoming quark and $\overline{{\left|\mathcal{M}\right|}^2}$
is the squared amplitude averaged (summed) over the spin and colour degrees of
freedom of the initial-state (final-state) particles.
The kinematically allowed ranges of $Q^2$ and $y$ are 
\begin{eqnarray}
Q_{\rm cut}^2& < &Q^2 < y_{\rm max}S-m_t^2,
\nonumber\\
\frac{Q^2+m_t^2}{S}& <& y < y_{\rm max},
\end{eqnarray}
where $Q_{\rm cut}^2$ defines the demarcation between photoproduction and
electroproduction and $y_{\rm max}$ is an experimental acceptance cut.
In Ref.~\cite{H1}, values for $Q_{\rm cut}^2$ and $y_{\rm max}$ are not
specified.
In our numerical analysis, we employ the typical values
$Q_{\rm cut}^2=4$~GeV$^2$ and $y_{\rm max}=0.95$.
Our numerical results are insensitive to the precise choice of $y_{\rm max}$
as long it is close to unity.
On the other hand, the dependence on $Q_{\rm cut}^2$ approximately cancels out
in the combination of photoproduction and electroproduction, as we explicitly
verify.

In the parton model of QCD, the hadronic cross section of
process~(\ref{eq:had}) is obtained by convoluting the partonic cross section
of process~(\ref{eq:sub}) with the appropriate PDF $F_q(\xi,\mu_F)$, where
$\mu_F$ is the factorisation scale, and summing over $q=u,c$, as
\begin{equation}
{\left(\frac{d\sigma}{dQ^2 dy}\right)}_{\rm hadr}
=\sum_{q=u,c}\int_0^1d\xi\, F_q(\xi,\mu_F)
 {\left(\frac{d\sigma}{dQ^2 dy}\right)}_{\rm part}.
\end{equation}
There are two candidate mass scales for $\mu_F$, namely $\sqrt{Q^2}$ and
$m_t$, and the optimal choice is likely to lie somewhere in between, at
$\mu_F=\left(\sqrt{Q^2}+m_t\right)/2$ say.
At any rate, we have $\mu_F\gg m_c$, so that charm is an active quark flavour
in the initial state, contributing at full strength via its PDF.

The expression for $\overline{{\left|\mathcal{M}\right|}^2}$ of
process~(\ref{eq:sub}) may be decomposed into a hadronic tensor $H^{\mu \nu}$
and a leptonic tensor $L^{\mu \nu}$, as
\begin{equation}
\label{Msquared}
\overline{{\left|\mathcal{M}\right|}^2}=\frac{e^2}{Q^4}L^{\mu\nu}H_{\mu\nu},
\end{equation}
where $e$ is the positron charge.
We have
\begin{equation}
H_{\mu\nu}=\frac{1}{2}\sum_{\rm spins}H_{\mu}^{\dagger}H_{\nu},
\end{equation}
where
\begin{eqnarray}
H_\mu&=&\overline{u}(p^\prime)
\left(F_1\gamma_\mu P_L+ F_2\gamma_{\mu}P_R+F_3p_\mu P_L
\right.\nonumber\\
&&{}+\left.F_4p_\mu P_R+F_5 p^{\prime}_\mu P_L+F_6 p_\mu^\prime P_R\right)u(p),
\end{eqnarray}
with helicity projectors $P_{L,R}=(1 \mp \gamma^5)/2$ and form factors
$F_1,\ldots,F_6$, which follow from the explicit evaluation of the Feynman
diagrams shown in Fig.~\ref{FeynmanSM} and their counterparts for an incoming
charm quark.
The leptonic tensor may be decomposed into transverse and longitudinal
components, as
\begin{equation}
L^{\mu\nu}=\frac{Q^2}{y^2}\left\{[1+(1-y)^2]\epsilon_T^{\mu\nu}
-4(1-y)\epsilon_L^{\mu\nu}\right\},
\end{equation}
where
\begin{eqnarray}
\epsilon_T^{\mu\nu}&=&-g^{\mu\nu}
+\frac{4Q^2}{\left(Q^2+m_t^2\right)^2}p^\mu p^\nu
+\frac{2}{Q^2+m_t^2}(p^\mu q^\nu +p^\nu q^\mu),
\nonumber\\
\epsilon_L^{\mu\nu}&=&
-\frac{1}{Q^2}\left(\frac{2Q^2}{Q^2+m_t^2}p^\mu+q^\mu\right)
\left(\frac{2Q^2}{Q^2+m_t^2}p^\nu+q^\nu\right).
\end{eqnarray}
To obtain the transversal and longitudinal parts of the cross section, the
hadron tensor is contracted with the transversal and longitudinal parts of the
lepton tensor, respectively.

We generate and evaluate the Feynman diagrams in Fig.~\ref{FeynmanSM}, with
the virtual-photon leg amputated, with the help of the program packages
{\it FeynArts} \cite{FAalt,FAFCSUSY} and {\it FormCalc} \cite{FAFCSUSY,FCLT}.
We work in 't~Hooft-Feynman gauge and use dimensional regularisation to
extract the ultraviolet (UV) divergences.
We perform the Passarino-Veltman reduction and calculate the squared amplitude
(\ref{Msquared}) using the program package {\it FeynCalc} \cite{FeynCalc}.
For the numerical evaluation of the standard scalar one-loop integrals, we
employ the program package {\it LoopTools} \cite{FCLT,LoopTools}.
We perform the numerical integration with the aid of the program package
{\it Cuba} \cite{Cuba}.
As a check, we also calculate the amplitude of process~(\ref{eq:sub}) with the
help of the program package {\it FeynCalc} and by hand.
All three independent calculations are found to lead to the same result.
Furthermore, we verify the cancellation of the UV divergences, current
conservation, and the reality of the squared amplitude. 

\subsection{Photoproduction}

In the photoproduction limit, the virtual photon is considered as real, with
$Q^2=0$, so that the longitudinal part of the cross section vanishes.
In turn, its energy distribution is described in the Weizs\"acker-Williams
approximation by the electron-to-photon splitting function
\begin{equation}
f_\gamma(y)=\frac{\alpha}{2\pi}\left[
\frac{1+(1-y)^2}{y}\ln\frac{Q_{\rm cut}^2}{Q_{\rm min}^2}
+2y m_e^2\left(\frac{1}{Q_{\rm cut}^2}-\frac{1}{Q_{\rm min}^2}\right)\right],
\end{equation}
where $\alpha=e^2/(4\pi)$ is Sommerfeld's fine-structure constant, $m_e$ is
the electron mass, and $Q_{\rm min}^2=y^2m_e^2/(1-y)$ corresponds to the
kinematic lower bound.
Thus, the cross section of process~(\ref{eq:sub}) in photoproduction is given
by
\begin{equation}
\left(\frac{d\sigma}{dy}\right)_{\rm part}=f_\gamma(y)\sigma_\gamma(y),
\end{equation}
with
\begin{equation}
\sigma_\gamma=\frac{\pi}{\xi yS}\overline{{\left|\mathcal{M}_\gamma\right|}^2}
\delta(\xi yS-m_t^2),
\end{equation}
where $\mathcal{M}_\gamma$ is the amplitude of $\gamma q\to t$.
The kinematically allowed range of $y$ is
\begin{equation}
\frac{m_t^2}{S} < y < y_{\rm max}.
\end{equation}
The hadronic cross section is again obtained by convoluting the partonic cross
section for a given incoming quark with the corresponding PDF and summing over
the incoming-quark flavours.
Here, the only candidate mass scale for $\mu_F$ is of order $m_t$, and it is
plausible to choose $\mu_F=m_t/2$ so that there is a smooth transition between
photoproduction and electroproduction.
Again, we have $\mu_F\gg m_c$, so that the use of a charm PDF is justified.

\section{Minimal and non-minimal flavour violation in the MSSM}
\label{sec:three}

\begin{center}
\begin{figure}
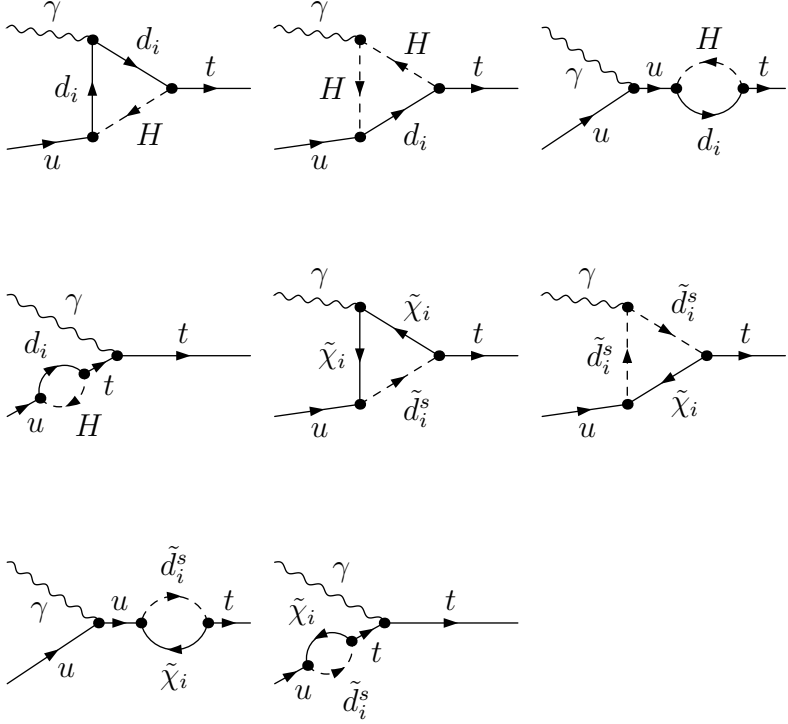


\unitlength=0.7bp%

\begin{feynartspicture}(432,450)(3,3)

\FADiagram{}
\FAProp(0.,15.)(7.,14.)(0.,){/Sine}{0}
\FALabel(3.7192,15.5544)[b]{$\gamma$}
\FAProp(0.,5.)(7.,6.)(0.,){/Straight}{1}
\FALabel(3.7192,4.44558)[t]{$u$}
\FAProp(20.,10.)(13.5,10.)(0.,){/Straight}{-1}
\FALabel(16.75,11.07)[b]{$t$}
\FAProp(7.,14.)(7.,6.)(0.,){/Straight}{-1}
\FALabel(5.93,10.)[r]{$d_i$}
\FAProp(7.,14.)(13.5,10.)(0.,){/Straight}{1}
\FALabel(10.5824,12.8401)[bl]{$d_i$}
\FAProp(7.,6.)(13.5,10.)(0.,){/ScalarDash}{-1}
\FALabel(10.5824,7.15993)[tl]{$H$}
\FAVert(7.,14.){0}
\FAVert(7.,6.){0}
\FAVert(13.5,10.){0}

\FADiagram{}
\FAProp(0.,15.)(7.,14.)(0.,){/Sine}{0}
\FALabel(3.7192,15.5544)[b]{$\gamma$}
\FAProp(0.,5.)(7.,6.)(0.,){/Straight}{1}
\FALabel(3.7192,4.44558)[t]{$u$}
\FAProp(20.,10.)(13.5,10.)(0.,){/Straight}{-1}
\FALabel(16.75,11.07)[b]{$t$}
\FAProp(7.,14.)(7.,6.)(0.,){/ScalarDash}{1}
\FALabel(5.93,10.)[r]{$H$}
\FAProp(7.,14.)(13.5,10.)(0.,){/ScalarDash}{-1}
\FALabel(10.5824,12.8401)[bl]{$H$}
\FAProp(7.,6.)(13.5,10.)(0.,){/Straight}{1}
\FALabel(10.5824,7.15993)[tl]{$d_i$}
\FAVert(7.,14.){0}
\FAVert(7.,6.){0}
\FAVert(13.5,10.){0}

\FADiagram{}
\FAProp(0.,15.)(7.5,10.)(0.,){/Sine}{0}
\FALabel(3.37021,11.6903)[tr]{$\gamma$}
\FAProp(0.,5.)(7.5,10.)(0.,){/Straight}{1}
\FALabel(4.12979,6.69032)[tl]{$u$}
\FAProp(20.,10.)(16.5,10.)(0.,){/Straight}{-1}
\FALabel(18.25,11.07)[b]{$t$}
\FAProp(7.5,10.)(11.,10.)(0.,){/Straight}{1}
\FALabel(9.25,11.07)[b]{$u$}
\FAProp(16.5,10.)(11.,10.)(-0.8,){/Straight}{-1}
\FALabel(13.75,6.73)[t]{$d_i$}
\FAProp(16.5,10.)(11.,10.)(0.8,){/ScalarDash}{1}
\FALabel(13.75,13.27)[b]{$H$}
\FAVert(7.5,10.){0}
\FAVert(16.5,10.){0}
\FAVert(11.,10.){0}

\FADiagram{}
\FAProp(0.,15.)(9.,10.)(0.,){/Sine}{0}
\FALabel(4.77275,13.3749)[bl]{$\gamma$}
\FAProp(0.,5.)(2.7,6.5)(0.,){/Straight}{1}
\FALabel(1.62275,4.87506)[tl]{$u$}
\FAProp(20.,10.)(9.,10.)(0.,){/Straight}{-1}
\FALabel(14.5,11.07)[b]{$t$}
\FAProp(9.,10.)(6.3,8.5)(0.,){/Straight}{-1}
\FALabel(7.92275,8.37506)[tl]{$t$}
\FAProp(2.7,6.5)(6.3,8.5)(-0.8,){/Straight}{1}
\FALabel(3.42725,9.81494)[br]{$d_i$}
\FAProp(2.7,6.5)(6.3,8.5)(0.8,){/ScalarDash}{-1}
\FALabel(5.57275,5.18506)[tl]{$H$}
\FAVert(9.,10.){0}
\FAVert(2.7,6.5){0}
\FAVert(6.3,8.5){0}

\FADiagram{}
\FAProp(0.,15.)(7.,14.)(0.,){/Sine}{0}
\FALabel(3.7192,15.5544)[b]{$\gamma$}
\FAProp(0.,5.)(7.,6.)(0.,){/Straight}{1}
\FALabel(3.7192,4.44558)[t]{$u$}
\FAProp(20.,10.)(13.5,10.)(0.,){/Straight}{-1}
\FALabel(16.75,11.07)[b]{$t$}
\FAProp(7.,14.)(7.,6.)(0.,){/Straight}{1}
\FALabel(5.93,10.)[r]{$\tilde \chi_i$}
\FAProp(7.,14.)(13.5,10.)(0.,){/Straight}{-1}
\FALabel(10.5824,12.8401)[bl]{$\tilde \chi_i$}
\FAProp(7.,6.)(13.5,10.)(0.,){/ScalarDash}{1}
\FALabel(10.5824,7.15993)[tl]{$\tilde d_i^s$}
\FAVert(7.,14.){0}
\FAVert(7.,6.){0}
\FAVert(13.5,10.){0}

\FADiagram{}
\FAProp(0.,15.)(7.,14.)(0.,){/Sine}{0}
\FALabel(3.7192,15.5544)[b]{$\gamma$}
\FAProp(0.,5.)(7.,6.)(0.,){/Straight}{1}
\FALabel(3.7192,4.44558)[t]{$u$}
\FAProp(20.,10.)(13.5,10.)(0.,){/Straight}{-1}
\FALabel(16.75,11.07)[b]{$t$}
\FAProp(7.,14.)(7.,6.)(0.,){/ScalarDash}{-1}
\FALabel(5.93,10.)[r]{$\tilde d_i^s$}
\FAProp(7.,14.)(13.5,10.)(0.,){/ScalarDash}{1}
\FALabel(10.5824,12.8401)[bl]{$\tilde d_i^s$}
\FAProp(7.,6.)(13.5,10.)(0.,){/Straight}{-1}
\FALabel(10.5824,7.15993)[tl]{$\tilde \chi_i$}
\FAVert(7.,14.){0}
\FAVert(7.,6.){0}
\FAVert(13.5,10.){0}

\FADiagram{}
\FAProp(0.,15.)(7.5,10.)(0.,){/Sine}{0}
\FALabel(3.37021,11.6903)[tr]{$\gamma$}
\FAProp(0.,5.)(7.5,10.)(0.,){/Straight}{1}
\FALabel(4.12979,6.69032)[tl]{$u$}
\FAProp(20.,10.)(16.5,10.)(0.,){/Straight}{-1}
\FALabel(18.25,11.07)[b]{$t$}
\FAProp(7.5,10.)(11.,10.)(0.,){/Straight}{1}
\FALabel(9.25,11.07)[b]{$u$}
\FAProp(16.5,10.)(11.,10.)(-0.8,){/Straight}{1}
\FALabel(13.75,6.73)[t]{$\tilde \chi_i$}
\FAProp(16.5,10.)(11.,10.)(0.8,){/ScalarDash}{-1}
\FALabel(13.75,13.27)[b]{$\tilde d_i^s$}
\FAVert(7.5,10.){0}
\FAVert(16.5,10.){0}
\FAVert(11.,10.){0}

\FADiagram{}
\FAProp(0.,15.)(9.,10.)(0.,){/Sine}{0}
\FALabel(4.77275,13.3749)[bl]{$\gamma$}
\FAProp(0.,5.)(2.7,6.5)(0.,){/Straight}{1}
\FALabel(1.62275,4.87506)[tl]{$u$}
\FAProp(20.,10.)(9.,10.)(0.,){/Straight}{-1}
\FALabel(14.5,11.07)[b]{$t$}
\FAProp(9.,10.)(6.3,8.5)(0.,){/Straight}{-1}
\FALabel(7.92275,8.37506)[tl]{$t$}
\FAProp(2.7,6.5)(6.3,8.5)(-0.8,){/Straight}{-1}
\FALabel(3.42725,9.81494)[br]{$\tilde \chi_i$}
\FAProp(2.7,6.5)(6.3,8.5)(0.8,){/ScalarDash}{1}
\FALabel(5.57275,5.18506)[tl]{$\tilde d_i^s$}
\FAVert(9.,10.){0}
\FAVert(2.7,6.5){0}
\FAVert(6.3,8.5){0}

\end{feynartspicture}

\caption{\label{FeynmanMFV}Additional Feynman diagrams of partonic
subprocess~(\ref{eq:sub}) arising in the MFV MSSM.
Here, $\tilde d_i^s$ with $i=1,2,3$ and $s=1,2$ are the down-type squarks, and
$\tilde\chi_i$ with $i=1,2$ are the charginos.}

\end{figure}
\end{center}

In the MSSM, there are two sources of FC phenomena \cite{Hollik}.
The first one is due to flavour mixing in the quark sector, just as in the SM.
It is produced by the different flavour rotations in the up- and down-quark
sectors, and its strength is driven by the off-diagonal CKM matrix elements.
This mixing produces FC electroweak (EW) interaction terms involving CCs, now
also involving charged Higgs bosons, and SUSY EW interaction terms of the
chargino-quark-squark type. 
Thus, the SM Feynman diagrams of Fig.~\ref{FeynmanSM} are supplemented by
those shown in Fig.~\ref{FeynmanMFV}, in which either charged Higgs bosons and
down-type quarks or charginos and down-type squarks circulate in the loops.
In the MFV MSSM, this is the only source of FC phenomena beyond the SM.

The second source of FC phenomena, which is present in the NMFV MSSM, is due
to the possible misalignment between the rotations that diagonalise the quark
and squark sectors.
When the squark mass matrix is expressed in the basis where the squark fields
are parallel to the quark fields (the super CKM basis), it is in general
non-diagonal in flavour space.
This quark-squark misalignment produces new FC terms in NC as well as in CC
interactions. 
In the SUSY QCD sector, the FC interaction terms involve NCs of the
gluino-quark-squark type.
In the case of process~(\ref{eq:sub}), this gives rise to the additional
Feynman diagrams shown in the first row of Fig.~\ref{FeynmanNMFV}.
In the SUSY EW sector, the FC interaction terms involve NCs of the
neutralino-quark-squark and the chargino-quark-squark type. 
The first type appears exclusively due to quark-squark misalignment, as in the
SUSY-QCD case, whereas the second type receives contributions from both
sources, quark-squark misalignment and CKM mixing.
The additional Feynman diagrams involving neutralino-quark-squark interactions
are displayed in the second row of Fig.~\ref{FeynmanNMFV}.

In order to simplify our analysis in the NMFV MSSM, we take the CKM matrix to
be diagonal, so that SM contributions (cf.\ Fig.~\ref{FeynmanSM}) and the
genuine MFV-MSSM contributions, i.e.\ the charged-Higgs-quark contributions
and the part of the chargino-squark contributions due to CKM mixing
(cf.\ Fig.~\ref{FeynmanMFV}), are zero.
We are then left with the gluino-squark and neutralino-squark contributions
(cf.\ Fig.~\ref{FeynmanNMFV}) and the residual chargino-squark contributions.

Furthermore, we assume that the non-CKM squark mixing is significant only for
transitions between the second- and third-generation squarks, and that there
is only left-left (LL) mixing, given by an ansatz similar as in
Ref.~\cite{AnsatzSher}, where it is proportional to the product of the
SUSY masses involved.
This assumption is theoretically well motivated by the flavour-off-diagonal
squark squared-mass entries that are radiatively induced via the evolution
from high energies down to the EW scale according to the renormalisation group
equations (RGEs) \cite{Hikasa}.
These RGEs predict that the FC LL entries scale with the square of the
soft-SUSY-breaking masses, in contrast with the left-right (LR) or right-left
(RL) and the right-right (RR) entries, which scale with one or zero powers,
respectively.
Thus, the hierarchy LL${}\gg{}$LR, RL${}\gg{}$RR is usually assumed.
The same estimates also indicate that the LL entry for the mixing between the
second- and third-generation squarks is the dominant one due to the larger
quark-mass factors involved.
On the other hand, the LR and RL entries are experimentally more constrained,
mainly by $b \to s \gamma$ data~\cite{ciuchini}.
With the previous assumption, the squark squared-mass matrices in the
$(\tilde{u}_L,\tilde{c}_L,\tilde{t}_L,\tilde{u}_R,\tilde{c}_R,\tilde{t}_R)$
and
$(\tilde{d}_L,\tilde{s}_L,\tilde{b}_L,\tilde{d}_R,\tilde{s}_R,\tilde{b}_R)$
bases can be written as follows:
\begin{eqnarray}
M_{\tilde{u}}^2&=& \left( \begin{array}{cccccc} 
M_{L,u}^2 &0&0& m_u X_u &0&0 \\
0& M_{L,c}^2 & \lambda_{LL}^t M_{L,c} M_{L,t} &0& m_c X_c &0 \\
0& \lambda_{LL}^t M_{L,c} M_{L,t} & M_{L,t}^2 &0&0& m_t X_t \\
m_u X_u &0&0& M_{R,u}^2 &0&0 \\
0& m_c X_c &0&0& M_{R,c}^2 &0 \\
0&0& m_t X_t &0&0& M_{R,t}^2
\end{array} \right), 
\label{eq.usquarkmass}\\
M_{\tilde{d}}^2&=& \left( \begin{array}{cccccc} 
M_{L,d}^2 &0&0& m_d X_d &0&0 \\
0& M_{L,s}^2 & \lambda_{LL}^b M_{L,s} M_{L,b} &0& m_s X_s &0 \\
0& \lambda_{LL}^b M_{L,s} M_{L,b} & M_{L,b}^2 &0&0& m_b X_b \\
m_d X_d &0&0& M_{R,d}^2 &0&0 \\
0& m_s X_s &0&0& M_{R,s}^2 &0 \\
0&0& m_b X_b &0&0& M_{R,b}^2
\end{array} \right),
\label{eq.dsquarkmass} 
\end{eqnarray}
where
\begin{eqnarray}
M_{L,q}^2&=& M_{\tilde{Q},q}^2 + m_q^2 
+ \cos(2\beta) M_Z^2 \left(T_3^q - Q_q s_w^2\right),
\nonumber\\
M_{R,q}^2&=&M_{\tilde{U},q}^2+m_q^2
+\cos(2\beta) Q_q s_w^2 M_Z^2 \qquad (q=u,c,t),
\nonumber\\
M_{R,q}^2&=&M_{\tilde{D},q}^2+m_q^2
+\cos(2\beta) Q_q s_w^2 M_Z^2 \qquad (q=d,s,b),
\nonumber\\
X_q&=&A_q-\mu (\tan\beta)^{-2 T_3^q},
\end{eqnarray}
and $\lambda_{LL}^t$ and $\lambda_{LL}^b$ measure the squark flavour mixing
strengths in the $\tilde t$--$\tilde c$ and $\tilde b$--$\tilde s$ sectors,
respectively.
As for the SM parameters, $m_q$, $T_3^q$, and $Q_q$ are the mass, weak
isospin, and electric charge of quark $q$;
$M_Z$ is the $Z$-boson mass;
and $s_w=\sin\theta_w$ is the sine of the weak mixing angle $\theta_w$. 
As for the MSSM parameters,
$\tan\beta=v_2/v_1$ is the ratio of the vacuum expectation values of the
two Higgs doublets;
$\mu$ is the Higgs-higgsino mass parameter;
$A_q$ are the trilinear Higgs-sfermion couplings;
and $M_{\tilde{Q},q}$, $M_{\tilde{U},q}$, and $M_{\tilde{D},q}$ are the scalar
masses.
Owing to SU(2)${}_L$ invariance, we have $M_{\tilde Q,u} = M_{\tilde Q,d}$,
$M_{\tilde Q,c} = M_{\tilde Q,s}$, and $M_{\tilde Q,t} = M_{\tilde Q,b}$.
Further MSSM input parameters include the EW gaugino masses $M_1$ and $M_2$,
the gluino mass $M_3$, and the mass $M_{A^0}$ of the CP-odd neutral Higgs
boson $A^0$.

In order to reduce the NMFV-MSSM parameter space, we make the following
simplifying assumptions.
We assume that the flavour mixing strengths in the $\tilde t$--$\tilde c$ and
$\tilde b$--$\tilde s$ sectors coincide and put
$\lambda=\lambda_{LL}^t=\lambda_{LL}^b$ for a simpler notation.
Obviously, the choice $\lambda=0$ represents the case of zero squark flavour
mixing.
We assume that the various trilinear Higgs-sfermion couplings coincide and
write $A_0=A_u=A_c=A_t=A_d=A_s=A_b$.
We assume that the scalar masses coincide thus defining the common SUSY mass
scale $M_0=M_{\tilde{Q},q}=M_{\tilde{U},\{u,c,t\}}=M_{\tilde{D},\{d,s,b\}}$.
As for the gaugino masses, we impose the GUT relation
$M_1=(5/3)(s_w^2/c_w^2)M_2$, where $c_w^2=1-s_w^2$, while we treat the gluino
mass parameter $M_3$ as independent.
We are thus left with eight independent MSSM parameters, namely,
$\tan\beta$, $M_{A^0}$, $M_0$, $M_2$, $M_3$, $A_0$, $\mu$, and $\lambda$.

\section{Numerical analysis} 
\label{sec:four}

In this section, we present our numerical results.
We adopt the SM parameters from Ref.~\cite{PDG} and the effective masses of
the down-type quarks from Ref.~\cite{Denner}:
\begin{eqnarray}
&&\alpha=1/137.035\,999\,11,\qquad \alpha_s=0.117\,6,
\nonumber\\
&&(\hbar c)^2=0.389\,379\,323~\mathrm{GeV^2 mb},\qquad
m_e=0.510\,998\,918~\mathrm{MeV},
\nonumber\\
&& M_W=80.403~\mathrm{GeV},\qquad M_Z=91.187\,6~\mathrm{GeV},\qquad
m_t=172.7~\mathrm{GeV},
\nonumber\\
&& m_d=0.041~\mathrm{GeV},\qquad m_s=0.15~\mathrm{GeV},\qquad
m_b=4.5~\mathrm{GeV}.
\end{eqnarray}
Here, $m_t$ and $m_b$ correspond to pole masses, while $m_d$ and $m_s$ were
determined so that their insertion in the perturbative formula for the
vacuum polarisation function of the photon reproduces the result extracted
from the total cross section of hadron production in $e^+e^-$ annihilation,
measured as a function of the CM energy, using a subtracted dispersion
relation.
This set of quark masses is especially appropriate for quantitative studies in
electroweak physics and is frequently employed in the literature.
It is used here for definiteness and convenience.
As a matter of principle, the precise definition of quark mass is not yet
fixed in an analysis of leading order (LO) in QCD like ours.
In fact, the freedom of choice of quark-mass definition contributes to the
theoretical uncertainty.
We employ the standard complex parametrisation of the CKM matrix in terms of
three angles $\theta_{12}$, $\theta_{23}$, $\theta_{13}$ and a phase
$\delta_{13}$,
\begin{equation}
V= \left( \begin{array}{ccc} 
c_{12} c_{13} & 
s_{12} c_{13} &
s_{13} e^{-i\delta_{13}} \\
-s_{12} c_{23} - c_{12} s_{23} s_{13} e^{i \delta_{13}} & 
c_{12} c_{23} - s_{12} s_{23} s_{13} e^{i \delta_{13}} &
s_{23} c_{13} \\
s_{12} s_{23} - c_{12} c_{23} s_{13} e^{i \delta_{13}} &
-c_{12} s_{23} - s_{12} c_{23} s_{13} e^{i \delta_{13}} &
c_{23} c_{13}
\end{array} \right),
\end{equation}
where $c_{ij}=\cos{\theta_{ij}}$ and $s_{ij}=\sin{\theta_{ij}}$, and
adopt from Ref.~\cite{PDG} the values
\begin{equation}
s_{12}=0.227\,2,\qquad s_{23}=0.042\,2,\qquad s_{13}=0.004\,0,\qquad
\delta_{13}=1.00.
\end{equation}
As for the proton PDFs, we employ the LO set CTEQ6L1 \cite{Cteq} by the
Coordinated Theoretical-Experimental Project on QCD (CTEQ) Collaboration.
We choose the factorisation scale to be $\mu_F=\left(\sqrt{Q^2}+m_t\right)/2$,
with the understanding that $Q^2=0$ in the case of photoproduction.
At HERA~II, electrons or positrons of energy $E_e=27.6$~GeV collide with
protons of energy $E_p=920$~GeV in the laboratory frame, yielding a CM energy
of $\sqrt{S}=319$~GeV.

\subsection{Standard Model}

\begin{center}
\begin{figure}
\epsfig{figure=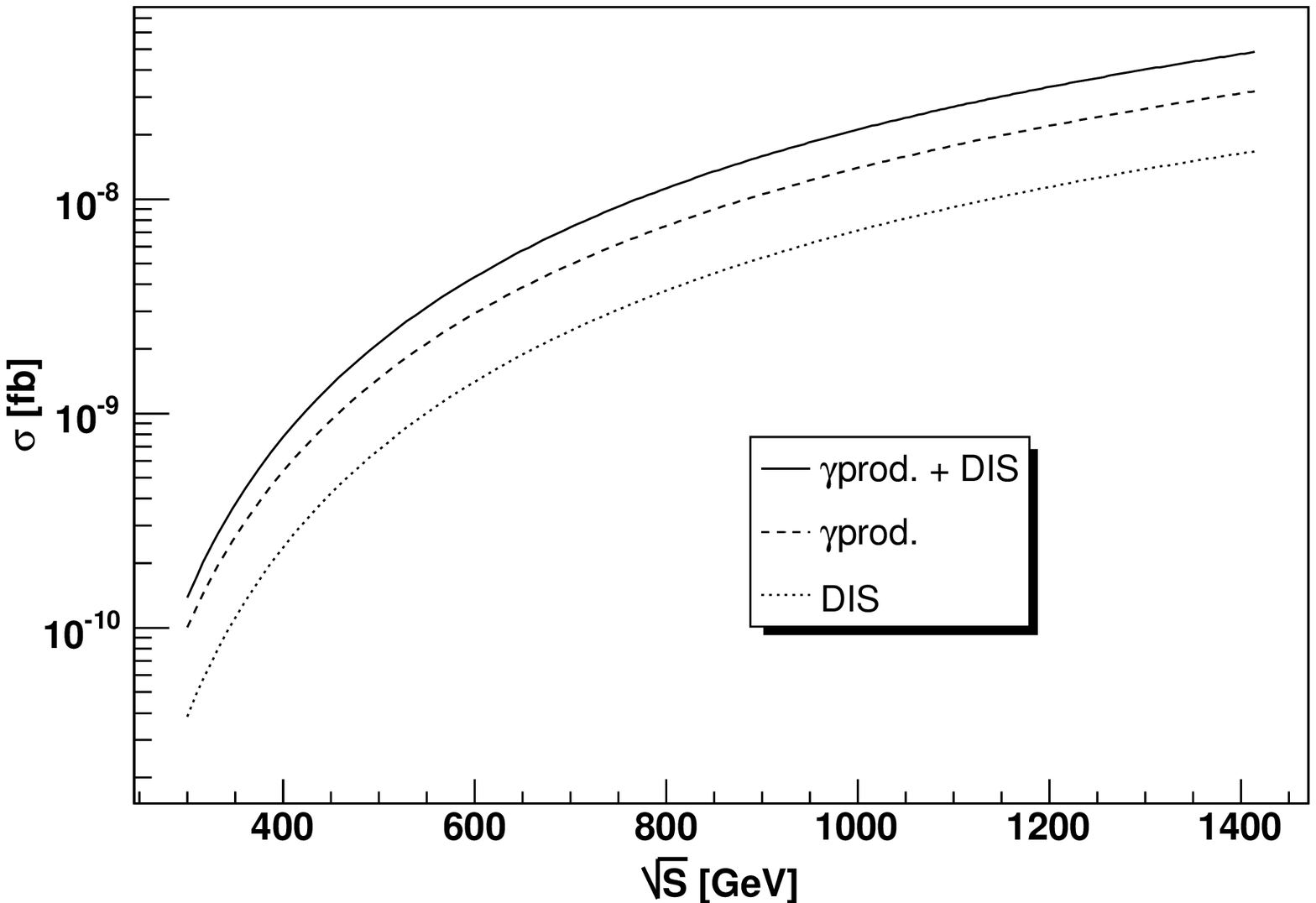,scale=0.7}
\caption{\label{SMSigS}Total cross section and its photoproduction and
electroproduction parts in the SM as functions of $\sqrt S$.}
\end{figure}
\end{center}

\begin{center}
\begin{figure}
\epsfig{figure=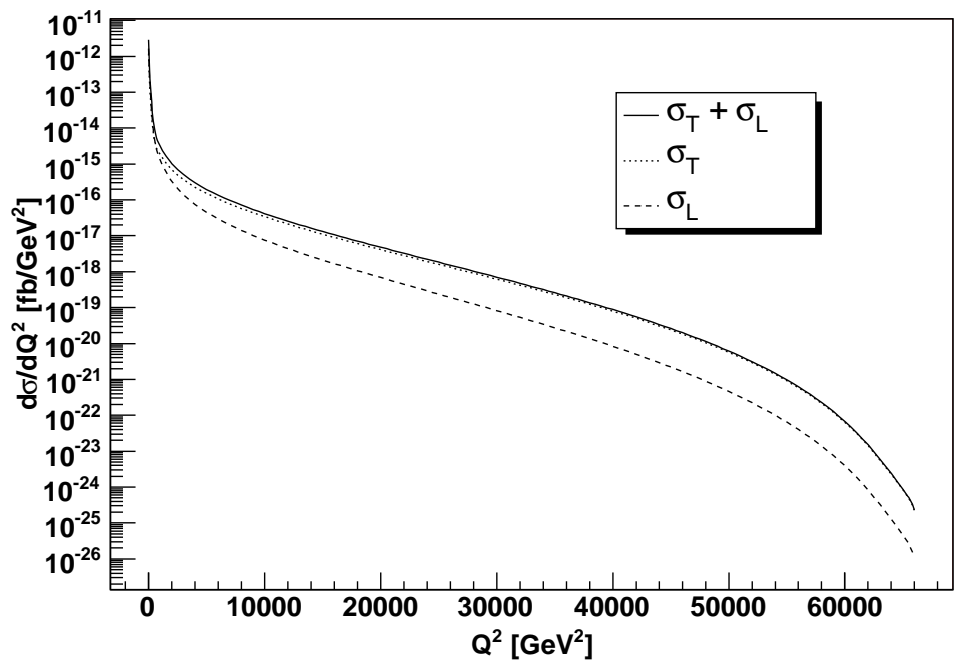,scale=0.7}
\caption{\label{SMdSigdQ}$Q^2$ distribution of the cross section and its
transversal and longitudinal parts in the SM under HERA experimental
conditions.}
\end{figure}
\end{center}

Figures~\ref{SMSigS} and \ref{SMdSigdQ} refer to the SM.
Figure~\ref{SMSigS} shows the total cross section of process~(\ref{eq:had})
as well as its photoproduction and electroproduction components as functions
of $\sqrt{S}$.
We observe that, for our choice of $Q_{\rm cut}^2$, the contribution due to
photoproduction  is approximately twice as large as the one due to
electroproduction.
At the CM energy of HERA, the cross section is of order $10^{-10}$~fb and thus
many orders of magnitude too small to be measurable.
A possible future electron-proton supercollider that uses the HERA proton
beam with energy $E_p=920$~GeV and the electron beam of the international
linear $e^+e^-$ collider (ILC) with energy $E_e=500$~GeV would have a CM
energy of 1357~GeV.
At this energy, the cross section is of order $10^{-8}$~fb and likewise not
measurable.  

Figure~\ref{SMdSigdQ} shows the $Q^2$ distribution of the electroproduction
cross section as well as its transversal and longitudinal parts.
We observe that the transversal part makes the major contribution. 

\subsection{MSSM with minimal flavour violation}

For definiteness, we assume that SUSY is broken according to the mSUGRA
scenario of a Grand Unified Theory (GUT).
We assign the following default values to the mSUGRA input parameters at the
GUT scale:
\begin{eqnarray}
&&\tan{\beta}=56,\qquad m_0=1.25~\mathrm{TeV},\qquad m_{1/2}=140~\mathrm{GeV},
\nonumber\\
&& A_0=-260~\mathrm{GeV},\qquad \mathrm{sign}(\mu)=+1,
\label{eq:msugra}
\end{eqnarray}
where $m_0$ is the universal scalar mass,
$m_{1/2}$ is the universal gaugino mass, and
$A_0$ is the universal trilinear Higgs-sfermion coupling.
These values approximately maximise the cross section of
process~(\ref{eq:had}) and are in accordance with the experimental bounds from
$b\to s\gamma$ decay and on the masses of the various SUSY particles \cite{PDG}.
We calculate the MSSM mass spectrum with the help of the program package
{\it SuSpect} \cite{SuSpect}.

\begin{center}
\begin{figure}
\epsfig{figure=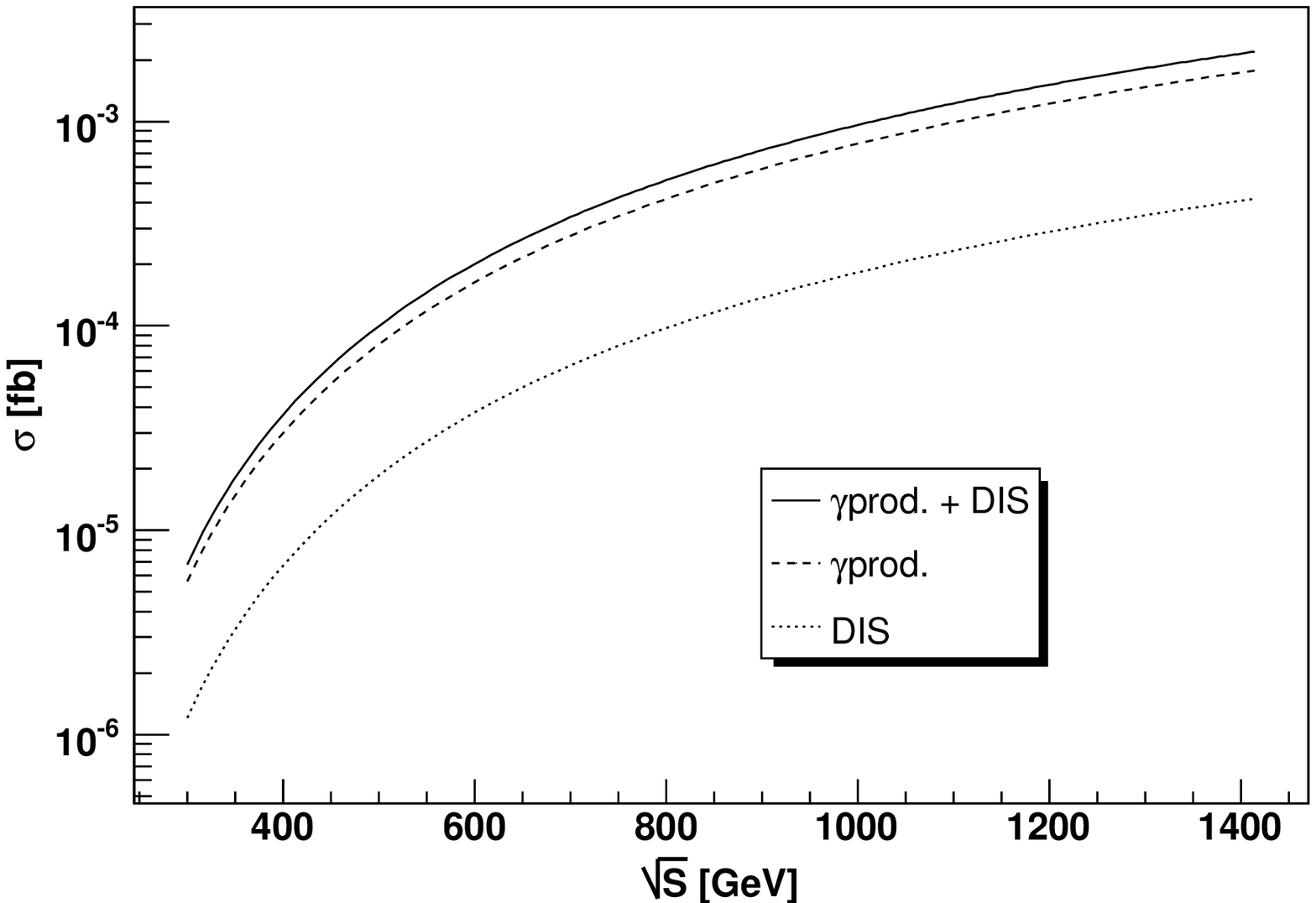,scale=0.7}
\caption{\label{MFVSigS}Total cross section and its photoproduction and
electroproduction parts in the MFV MSSM as functions of $\sqrt S$.}
\end{figure}
\end{center}

\begin{center}
\begin{figure}
\epsfig{figure=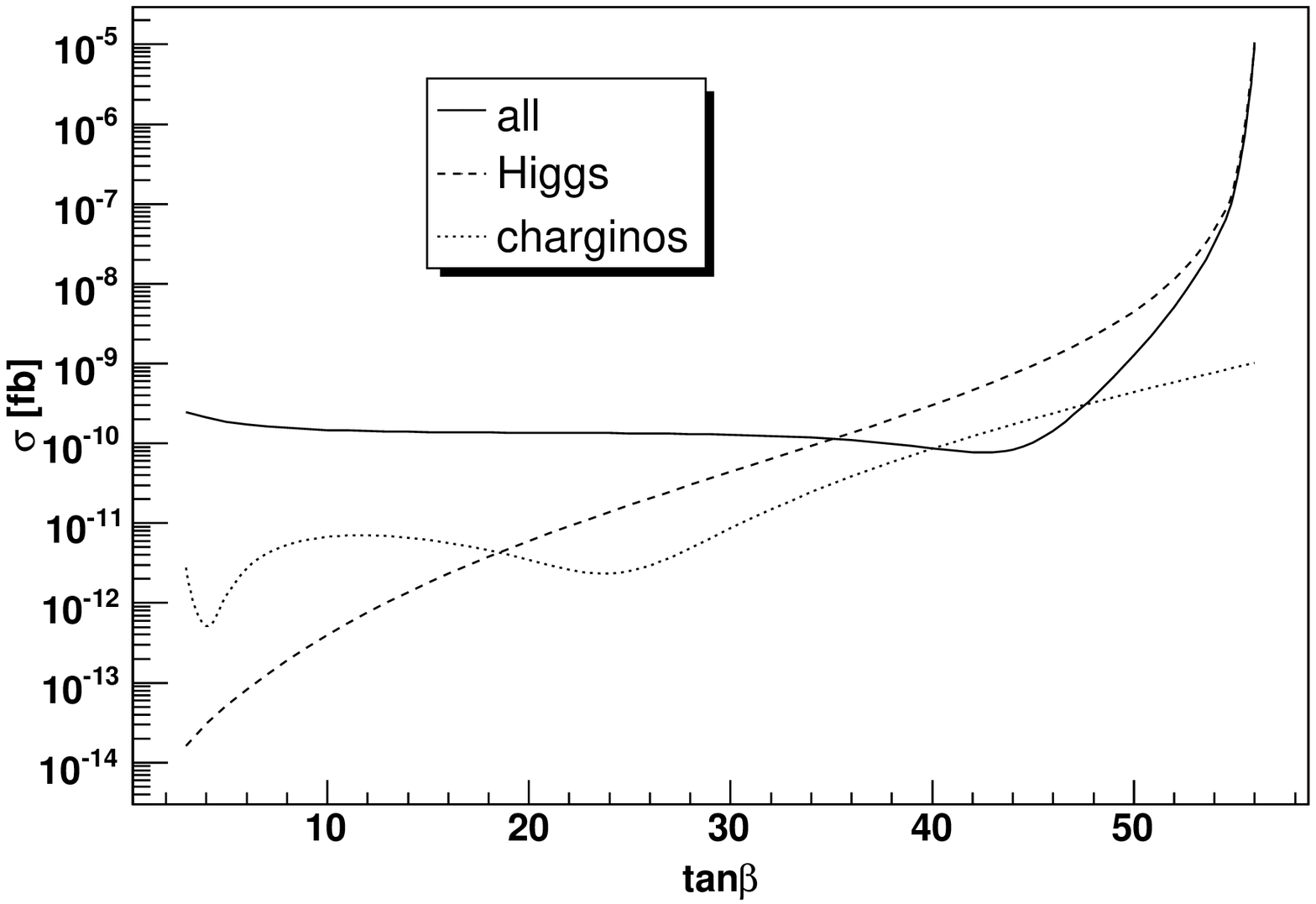,scale=0.7}
\caption{\label{MFVSigtb}Total cross section and its Higgs and chargino parts
in the MFV MSSM as functions of $\tan\beta$ under HERA experimental
conditions.}
\end{figure}
\end{center}

\begin{center}
\begin{figure}
\epsfig{figure=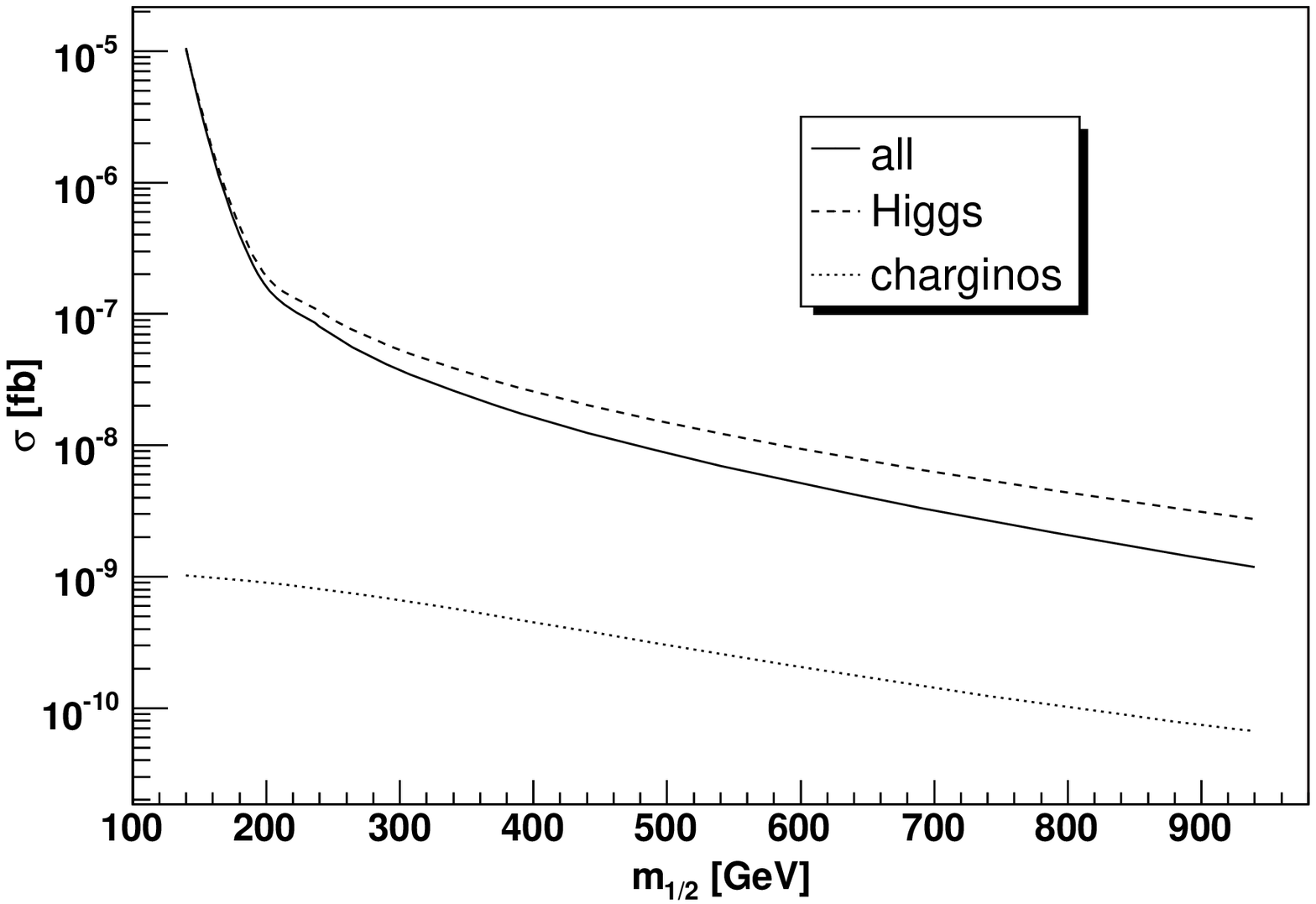,scale=0.7}
\caption{\label{MFVSigmhalf}Total cross section and its Higgs and chargino
parts in the MFV MSSM as functions of $m_{1/2}$ under HERA experimental
conditions.}
\end{figure}
\end{center}

\begin{center}
\begin{figure}
\epsfig{figure=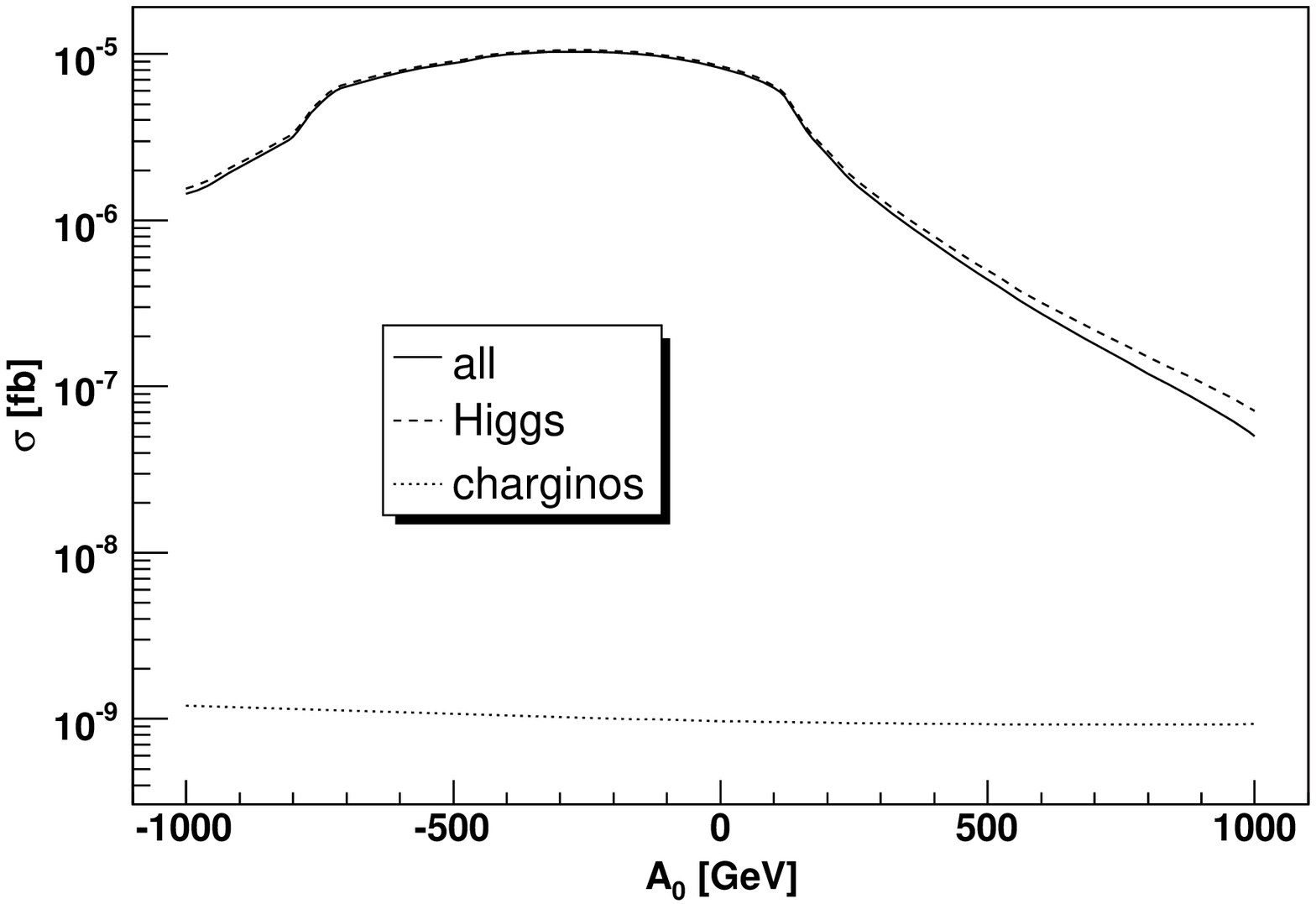,scale=0.7}
\caption{\label{MFVSigA0}Total cross section and its Higgs and chargino
parts in the MFV MSSM as functions of $A_0$ under HERA experimental
conditions.}
\end{figure}
\end{center}

Figures~\ref{MFVSigS}--\ref{MFVSigA0} all show the total cross section of
process~(\ref{eq:had}) in the MFV MSSM.
Specifically, Fig.~\ref{MFVSigS} displays the $\sqrt S$ dependence, also
separately for the photoproduction and electroproduction contributions, while
Figs.~\ref{MFVSigtb}--\ref{MFVSigA0} exhibit, for HERA experimental conditions,
the dependencies on $\tan\beta$, $m_{1/2}$, and $A_0$, respectively, also
separately for the charged-Higgs-boson and chargino contributions.

From Fig.~\ref{MFVSigS} we observe that the cross section is of order
$10^{-5}$~fb at HERA energy and of order $10^{-3}$~fb for the future
electron-proton supercollider mentioned above.
Both values are too small to yield measurable results. 

From Fig.~\ref{MFVSigtb} we learn that, as $\tan{\beta}$ approaches its upper
limit, the cross section strongly increases and is mainly generated by the
loop diagrams involving charged Higgs bosons.
For small values of $\tan{\beta}$, the SM contribution ($\approx 10^{-10}$~fb)
is dominant.
Negative interference effects between the charged-Higgs, chargino, and SM
contributions can be seen at large values of $\tan{\beta}$.

From Fig.~\ref{MFVSigmhalf} we see that, as $m_{1/2}$ approaches its lower
limit, the cross section strongly increases and is essentially made up by the
charged-Higgs-boson contribution alone.
The latter is dominant throughout the whole $m_{1/2}$ range, but there are
negative interference effects for all values of $m_{1/2}$.
The significant suppression of the chargino contribution for $\tan\beta=56$
familiar from Fig.~\ref{MFVSigtb} is actually present for all values of
$m_{1/2}$.

As is evident from  Fig.~\ref{MFVSigA0}, the cross section is largest for
$A_0=-260$~GeV and falls off by one (two) orders of magnitude as $A_0$ reaches
$-1$~TeV (1~TeV).
However, the variation with $A_0$ is less significant than those with
$\tan{\beta}$ and $m_{1/2}$.
For $\tan\beta=56$, the chargino contribution is several orders of magnitude
smaller than the charged-Higgs one and almost independent of $A_0$. 

We conclude that, in the MFV MSSM, the mSUGRA scenario characterised by the
input parameter values specified in Eq.~(\ref{eq:msugra}) approximately
maximises the cross section of process~(\ref{eq:had}), which still comes out
much below the threshold of observability at HERA and a future electron-proton
supercollider.

\subsection{MSSM with non-minimal flavour violation}

Prior to presenting our NMFV-MSSM results, we explain our choice of input
parameters.
Scanning the eight-dimensional parameter space defined at the end of
Section~\ref{sec:three}, we find that the following assignments, which we
henceforth take as default, approximately maximise the cross section of
process~(\ref{eq:had}):
\begin{eqnarray}
&&\tan\beta=8,\qquad
M_{A^0}=170~\mathrm{GeV},\qquad
M_0=475~\mathrm{GeV},\qquad
M_2=655~\mathrm{GeV},
\nonumber\\
&&M_3=195~\mathrm{GeV},\qquad
A_0=950~\mathrm{GeV},\qquad
\mu=345~\mathrm{GeV},\qquad
\lambda=0.73.
\end{eqnarray}
These parameters are in accordance with the lower bounds on the squark masses
of 100~GeV and with the experimental lower bounds for the masses of the other
SUSY particles \cite{PDG}.
They are also in accordance with the experimental bounds from $b\to s\gamma$
decay \cite{PDG}.
We calculate the MSSM spectrum using the program package {\it FeynHiggs}
\cite{FeynHiggs}.

\begin{center}
\begin{figure}
\epsfig{figure=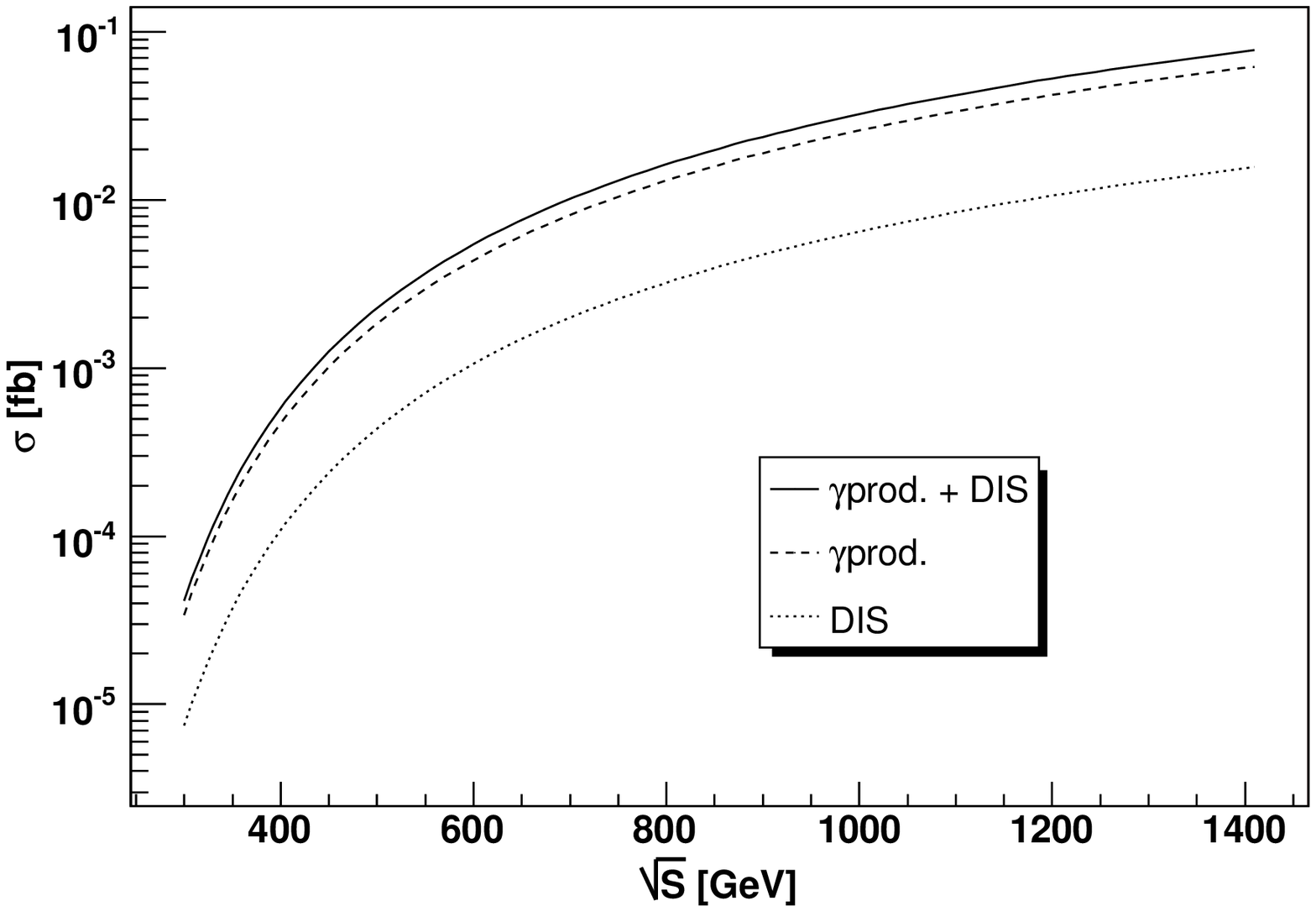,scale=0.7}
\caption{\label{NMFVSigS}Total cross section and its photoproduction and
electroproduction parts in the NMFV MSSM as functions of $\sqrt S$.}
\end{figure}
\end{center}

\begin{center}
\begin{figure}
\epsfig{figure=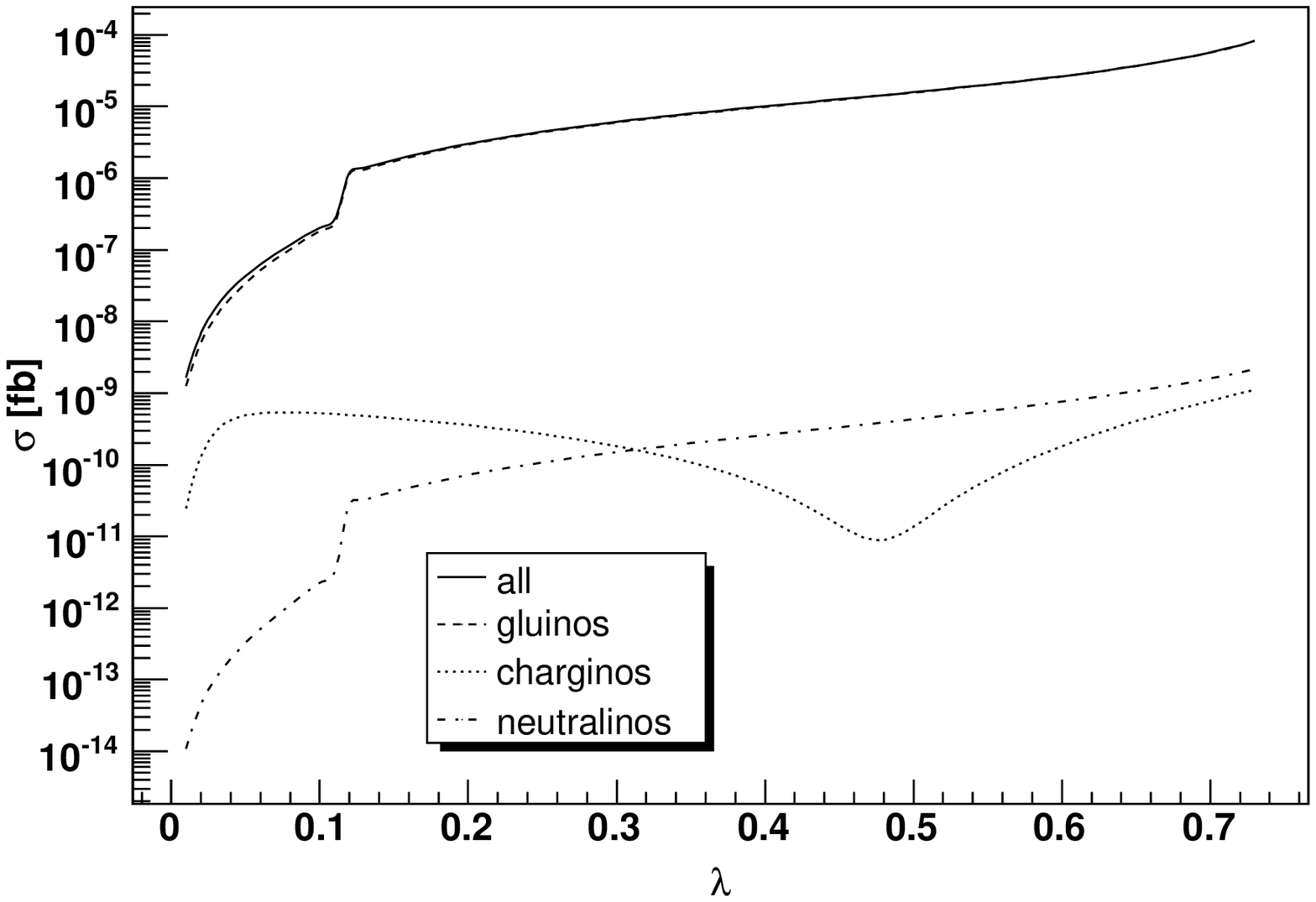,scale=0.7}
\caption{\label{NMFVSigLambda}Total cross section and its gluino, chargino,
and neutralino parts in the NMFV MSSM as functions of $\lambda$ under HERA
experimental conditions.}
\end{figure}
\end{center}

\begin{center}
\begin{figure}
\epsfig{figure=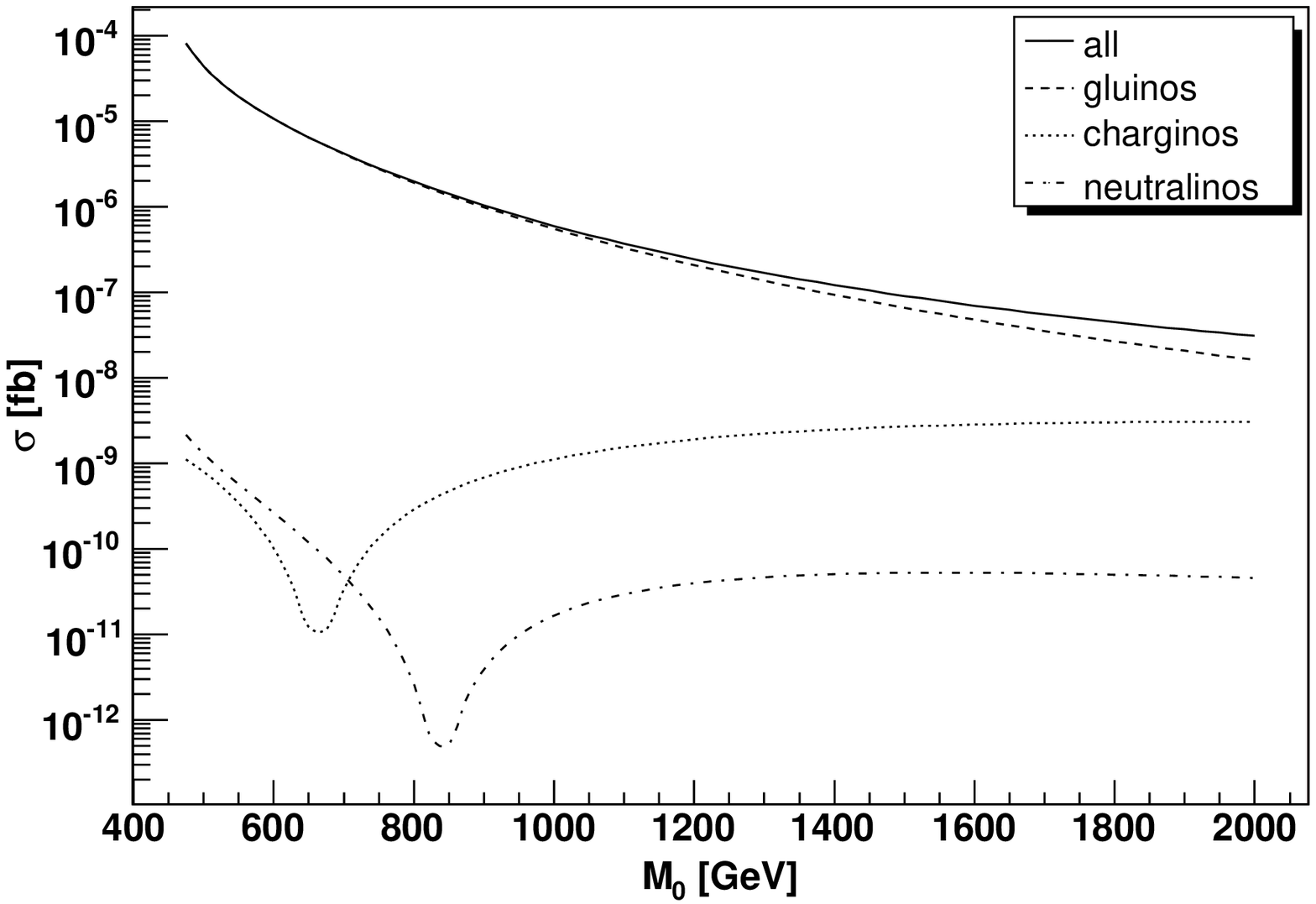,scale=0.7}
\caption{\label{NMFVSigMSUSY}Total cross section and its gluino, chargino,
and neutralino parts in the NMFV MSSM as functions of $M_0$ under HERA
experimental conditions.}
\end{figure}
\end{center}

\begin{center}
\begin{figure}
\epsfig{figure=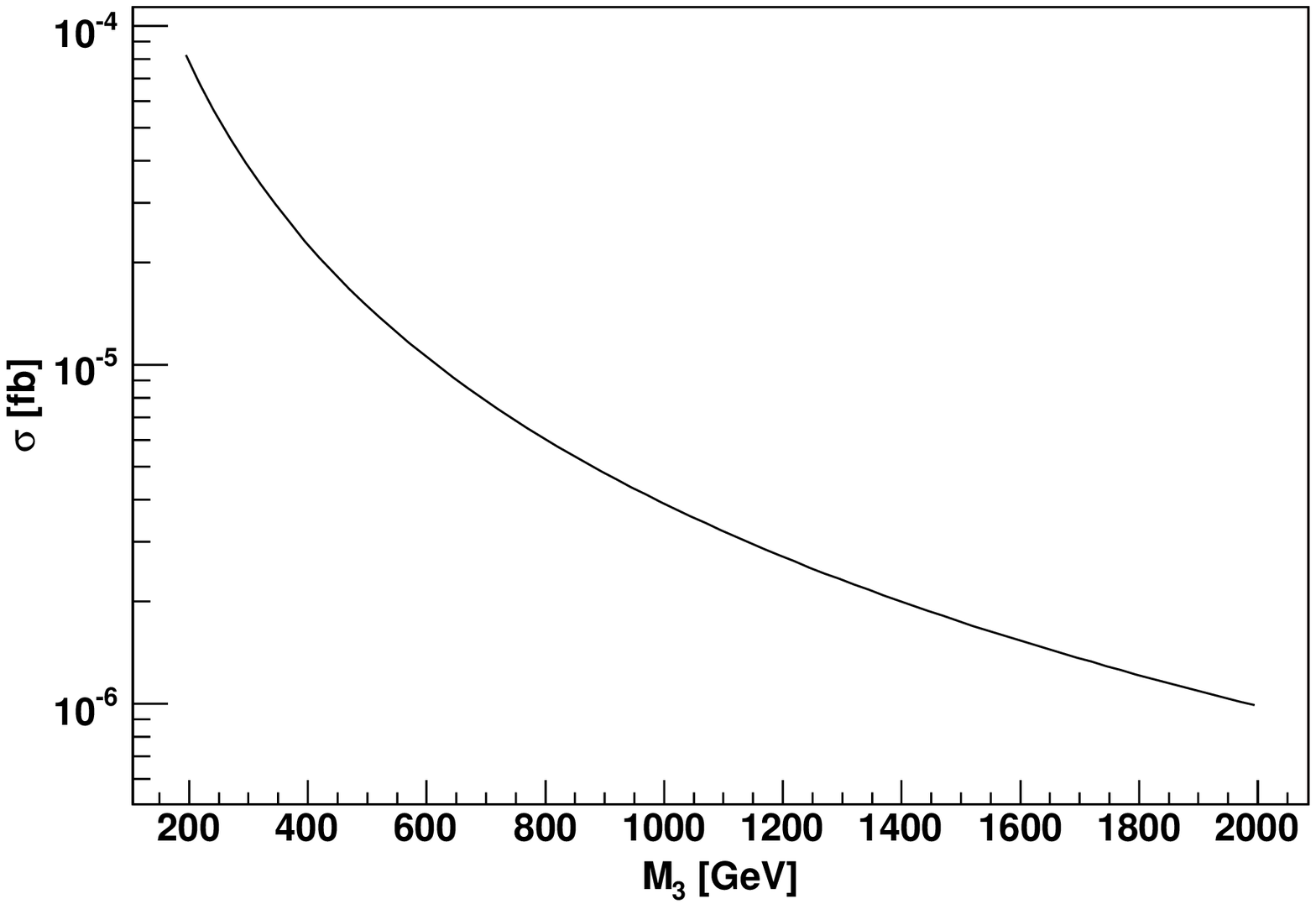,scale=0.7}
\caption{\label{NMFVSigMGl}Total cross section in the NMFV MSSM as function of
$M_3$ under HERA experimental conditions.}
\end{figure}
\end{center}

Figures~\ref{NMFVSigS}--\ref{NMFVSigMGl} all show the total cross section of
process~(\ref{eq:had}) in the NMFV MSSM.
Specifically, Fig.~\ref{NMFVSigS} displays the $\sqrt S$ dependence, also
separately for the photoproduction and electroproduction contributions, while
Figs.~\ref{NMFVSigLambda}--\ref{NMFVSigMGl} exhibit, for HERA experimental
conditions, the dependencies on $\lambda$, $M_0$, and $M_3$, respectively.
In Figs.~\ref{NMFVSigLambda} and \ref{NMFVSigMSUSY}, also the contributions
from loops involving gluinos, charginos, and neutralinos are shown separately.

From Fig.~\ref{NMFVSigS} we read off values of order $10^{-4}$~fb and
$10^{-1}$~fb for the cross sections at HERA and the future electron-proton
supercollider mentioned above, respectively, which is discouraging in the case
of HERA and challenging for the electron-proton supercollider, depending on
its luminosity.

From Fig.~\ref{NMFVSigLambda} we observe that the cross section strongly
increases with $\lambda$, by five orders of magnitude as $\lambda$ runs from 0
to 0.73.
For $\lambda>0.73$, the mass of the lightest squark is less than the lower
limit of 100~GeV.
The cross section is almost exhausted by the gluino contribution.
This may be understood by observing that the corresponding Feynman diagrams
are enhanced by a factor of $\alpha_s/\alpha$ relative to those of the
chargino and neutralino contributions.
The neutralino contribution exhibits a $\lambda$ dependence similar to the
full cross section, but is more than four orders of magnitude smaller.
The chargino contribution oscillates about a mean value of $10^{-10}$~fb in
the $\lambda$ range considered.
For $\lambda\alt0.3$ ($\lambda\agt0.3$), it overshoots (undershoots) the
neutralino contribution.

From Fig.~\ref{NMFVSigMSUSY} we observe that the cross section strongly
decreases with increasing value of $M_0$, by more than three orders of
magnitude as $M_0$ runs from 475~GeV to 2~TeV.
The dominant role of the gluino contribution observed in
Fig.~\ref{NMFVSigLambda} attenuates in the large-$M_0$ regime, where the
chargino contribution gains influence.
Nevertheless, there is a clear hierarchy among the gluino, chargino, and
neutralino contributions for $M_0\agt700$~GeV, the latter one being least
important.
For $M_0\alt700$~GeV, the neutralino contribution exceeds the chargino one.
The chargino and neutralino contributions exhibit minima at 650~GeV and
850~GeV, respectively.

From Fig.~\ref{NMFVSigMGl} we learn that the cross section decreases by about
two orders of magnitude as $M_3$ runs from 195~GeV, the Tevatron search limit
\cite{PDG}, to 2~TeV.

\section{Conclusion}
\label{sec:five}

In this paper, the photoproduction and photonic electroproduction of single
top quarks in electron-proton scattering was analysed for the first time.
The analysis was performed at one loop in the SM as well as the MSSM with
minimal and non-minimal flavour mixing.
In all three models, the cross section turned out to be too small to be
measurable at HERA.
The physics at HERA remains interesting because, in the light of our results,
the single-top-quark-like events seen by the H1 Collaboration might be a sign
of physics not only beyond the SM, but also beyond the MSSM with conserved R
parity.

\section*{Acknowledgements}

We are grateful to John Collins for useful communications regarding
Ref.~\cite{col}.
The work of B.A.K. was supported in part by the German Research Foundation DFG
through the Collaborative Research Center No.\ 676 {\it Particles, Strings and
the Early Universe---the Structure of Matter and Space-Time} and by the German
Federal Ministry for Education and Research BMBF through Grant No.\ 05~HT6GUA.

\newpage

\begin{appendix}

\begin{center}
\begin{figure}
\SetScale{0.8}
\unitlength=0.8bp%

  \begin{picture}(455,406) (15,-73)

    \SetWidth{0.5}

    \SetColor{Black}

    \Line(60,16)(180,16)

    \Line(210,16)(270,-44)

    \ArrowLine(315,121)(435,121)

    \ArrowLine(60,211)(195,211)

    \Photon(195,211)(300,136){7.5}{6}

    \ArrowLine(195,211)(315,316)

    \Text(121,220)[lb]{\Large{\Black{$k$}}}

    \Text(330,322)[lb]{\Large{\Black{$e^-$}}}

    \Text(25,205)[lb]{\Large{\Black{$e^-$}}}

    \Text(240,273)[lb]{\Large{\Black{$k^{\prime}$}}}

    \Text(230,142)[lb]{\Large{\Black{$\gamma$}}}

    \Text(270,173)[lb]{\Large{\Black{$q=k-k^\prime$}}}

    \Text(215,76)[lb]{\Large{\Black{$u, c$}}}

    \Text(250,48)[lb]{\Large{\Black{$p$}}}

    \Text(450,115)[lb]{\Large{\Black{$t$}}}

    \Text(374,130)[lb]{\Large{\Black{$p^\prime$}}}

    \Text(30,12)[lb]{\Large{\Black{$P$}}}

    \Text(104,40)[lb]{\Large{\Black{$P$}}}

    \Text(270,-65)[lb]{\Large{\Black{$X$}}}

    \Vertex(195,211){2.83}

    \Line(195,16)(255,-44)

    \Line(195,16)(300,121)

    \Line(60,1)(195,1)

    \Line(60,31)(195,31)

    \Line(195,16)(300,121)

    \GOval(195,16)(30,15)(0){0.882}

    \GOval(300,121)(15,15)(0){0.882}

  \end{picture}

\caption{\label{overview}Schematic representation of the hadronic process
(\ref{eq:had}) explaining the four-momentum assignments.}

\end{figure}
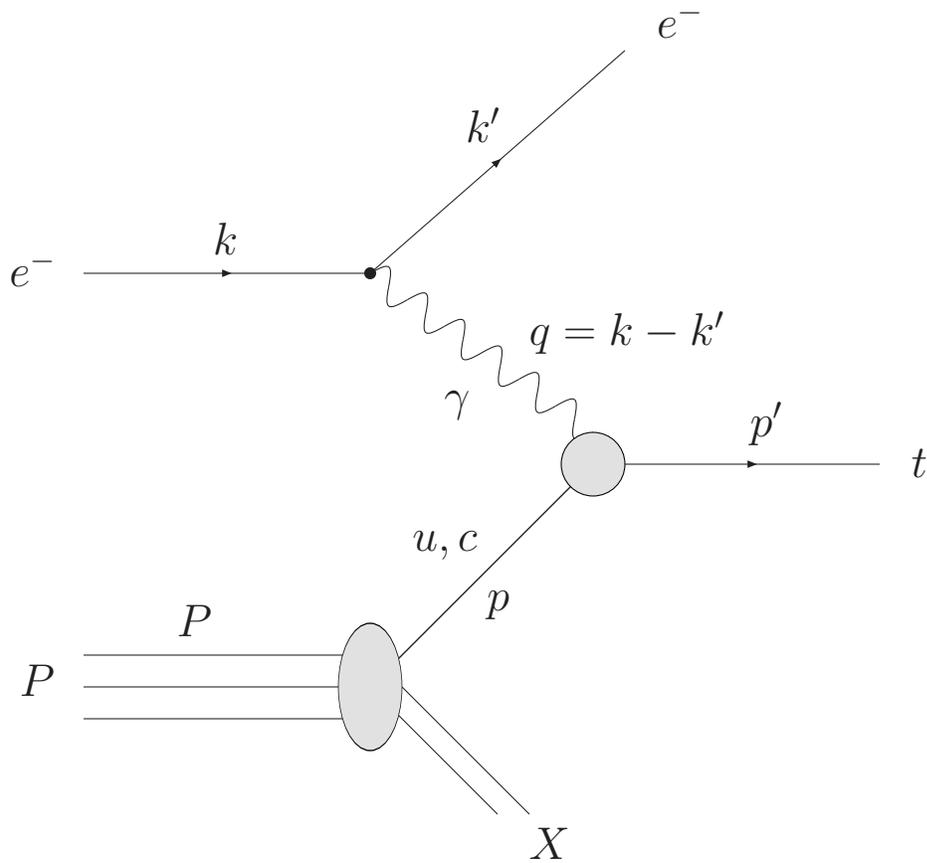
\end{center}


\begin{center}
\begin{figure}

\unitlength=0.7bp%

\begin{feynartspicture}(432,300)(3,2)

\FADiagram{}
\FAProp(0.,15.)(7.,14.)(0.,){/Sine}{0}
\FALabel(3.7192,15.5544)[b]{$\gamma$}
\FAProp(0.,5.)(7.,6.)(0.,){/Straight}{1}
\FALabel(3.7192,4.44558)[t]{$c$}
\FAProp(20.,10.)(13.5,10.)(0.,){/Straight}{-1}
\FALabel(16.75,11.07)[b]{$t$}
\FAProp(7.,14.)(7.,6.)(0.,){/ScalarDash}{-1}
\FALabel(5.93,10.)[r]{$\tilde u_a$}
\FAProp(7.,14.)(13.5,10.)(0.,){/ScalarDash}{1}
\FALabel(10.5824,12.8401)[bl]{$\tilde u_a$}
\FAProp(7.,6.)(13.5,10.)(0.,){/Straight}{0}
\FALabel(10.4513,7.37284)[tl]{$\tilde g$}
\FAVert(7.,14.){0}
\FAVert(7.,6.){0}
\FAVert(13.5,10.){0}

\FADiagram{}
\FAProp(0.,15.)(7.5,10.)(0.,){/Sine}{0}
\FALabel(3.37021,11.6903)[tr]{$\gamma$}
\FAProp(0.,5.)(7.5,10.)(0.,){/Straight}{1}
\FALabel(4.12979,6.69032)[tl]{$c$}
\FAProp(20.,10.)(16.5,10.)(0.,){/Straight}{-1}
\FALabel(18.25,11.07)[b]{$t$}
\FAProp(7.5,10.)(11.,10.)(0.,){/Straight}{1}
\FALabel(9.25,11.07)[b]{$u$}
\FAProp(16.5,10.)(11.,10.)(-0.8,){/Straight}{0}
\FALabel(13.75,6.98)[t]{$\tilde g$}
\FAProp(16.5,10.)(11.,10.)(0.8,){/ScalarDash}{-1}
\FALabel(13.75,13.27)[b]{$\tilde u_a$}
\FAVert(7.5,10.){0}
\FAVert(16.5,10.){0}
\FAVert(11.,10.){0}

\FADiagram{}
\FAProp(0.,15.)(9.,10.)(0.,){/Sine}{0}
\FALabel(4.77275,13.3749)[bl]{$\gamma$}
\FAProp(0.,5.)(2.7,6.5)(0.,){/Straight}{1}
\FALabel(1.62275,4.87506)[tl]{$c$}
\FAProp(20.,10.)(9.,10.)(0.,){/Straight}{-1}
\FALabel(14.5,11.07)[b]{$t$}
\FAProp(9.,10.)(6.3,8.5)(0.,){/Straight}{-1}
\FALabel(7.92275,8.37506)[tl]{$t$}
\FAProp(2.7,6.5)(6.3,8.5)(-0.8,){/Straight}{0}
\FALabel(3.54866,9.5964)[br]{$\tilde g$}
\FAProp(2.7,6.5)(6.3,8.5)(0.8,){/ScalarDash}{1}
\FALabel(5.57275,5.18506)[tl]{$\tilde u_a$}
\FAVert(9.,10.){0}
\FAVert(2.7,6.5){0}
\FAVert(6.3,8.5){0}

\FADiagram{}
\FAProp(0.,15.)(7.,14.)(0.,){/Sine}{0}
\FALabel(3.7192,15.5544)[b]{$\gamma$}
\FAProp(0.,5.)(7.,6.)(0.,){/Straight}{1}
\FALabel(3.7192,4.44558)[t]{$c$}
\FAProp(20.,10.)(13.5,10.)(0.,){/Straight}{-1}
\FALabel(16.75,11.07)[b]{$t$}
\FAProp(7.,14.)(7.,6.)(0.,){/ScalarDash}{-1}
\FALabel(5.93,10.)[r]{$\tilde u_a$}
\FAProp(7.,14.)(13.5,10.)(0.,){/ScalarDash}{1}
\FALabel(10.5824,12.8401)[bl]{$\tilde u_a$}
\FAProp(7.,6.)(13.5,10.)(0.,){/Straight}{0}
\FALabel(10.4513,7.37284)[tl]{$\tilde \chi_i^0$}
\FAVert(7.,14.){0}
\FAVert(7.,6.){0}
\FAVert(13.5,10.){0}

\FADiagram{}
\FAProp(0.,15.)(7.5,10.)(0.,){/Sine}{0}
\FALabel(3.37021,11.6903)[tr]{$\gamma$}
\FAProp(0.,5.)(7.5,10.)(0.,){/Straight}{1}
\FALabel(4.12979,6.69032)[tl]{$c$}
\FAProp(20.,10.)(16.5,10.)(0.,){/Straight}{-1}
\FALabel(18.25,11.07)[b]{$t$}
\FAProp(7.5,10.)(11.,10.)(0.,){/Straight}{1}
\FALabel(9.25,11.07)[b]{$u$}
\FAProp(16.5,10.)(11.,10.)(-0.8,){/Straight}{0}
\FALabel(13.75,6.98)[t]{$\tilde \chi_i^0$}
\FAProp(16.5,10.)(11.,10.)(0.8,){/ScalarDash}{-1}
\FALabel(13.75,13.27)[b]{$\tilde u_a$}
\FAVert(7.5,10.){0}
\FAVert(16.5,10.){0}
\FAVert(11.,10.){0}

\FADiagram{}
\FAProp(0.,15.)(9.,10.)(0.,){/Sine}{0}
\FALabel(4.77275,13.3749)[bl]{$\gamma$}
\FAProp(0.,5.)(2.7,6.5)(0.,){/Straight}{1}
\FALabel(1.62275,4.87506)[tl]{$c$}
\FAProp(20.,10.)(9.,10.)(0.,){/Straight}{-1}
\FALabel(14.5,11.07)[b]{$t$}
\FAProp(9.,10.)(6.3,8.5)(0.,){/Straight}{-1}
\FALabel(7.92275,8.37506)[tl]{$t$}
\FAProp(2.7,6.5)(6.3,8.5)(-0.8,){/Straight}{0}
\FALabel(3.54866,9.5964)[br]{$\tilde \chi_i^0$}
\FAProp(2.7,6.5)(6.3,8.5)(0.8,){/ScalarDash}{1}
\FALabel(5.57275,5.18506)[tl]{$\tilde u_a$}
\FAVert(9.,10.){0}
\FAVert(2.7,6.5){0}
\FAVert(6.3,8.5){0}

\end{feynartspicture}

\caption{\label{FeynmanNMFV}Additional Feynman diagrams of partonic
subprocess~(\ref{eq:sub}) arising in the NMFV MSSM.
Here, $\tilde g$ is the gluino, $\tilde u_a$ with $a=1,\ldots,6$ are the
up-type squarks, and $\tilde\chi_i^0$ with $i=1,\ldots,4$ are the
neutralinos.}

\end{figure}
\end{center}


\end{appendix}


\begin{thebibliography}{99}

\bibitem{H1}H1 Collaboration, A. Aktas et al.,
Eur.\ Phys.\ J. C 33 (2004) 9;\\
D.M. South, on behalf of the H1 Collaboration,
in: Proceedings of the 14th International Workshop on Deep Inelastic
Scattering (DIS 2006), Tsukuba, Japan, 20--24 April 2006 
(World Scientific, Singapore, 2007) p.~325.

\bibitem{ZEUS}ZEUS Collaboration, S. Chekanov et al.,
Phys.\ Lett.\ B 559 (2003) 153;\\
M. Corradi, on behalf of the ZEUS Collaboration,
in: Proceedings of the 14th International Workshop on Deep Inelastic
Scattering (DIS 2006), Tsukuba, Japan, 20--24 April 2006
(World Scientific, Singapore, 2007) p.~321.

\bibitem{CC}T. Stelzer, Z. Sullivan, S. Willenbrock,
Phys.\ Rev.\ D 56 (1997) 5919;\\
S. Moretti, K. Odagiri,
Phys.\ Rev.\ D 57 (1998) 3040.

\bibitem{K0D0}F. Gabbiani, E. Gabrielli, A. Masiero, L.Silvestrini,
Nucl.\ Phys.\ B 477 (1996) 321;\\
M. Misiak, S. Pokorski, J.Rosiek,
Adv.\ Ser.\ Direct.\ High Energy Phys.\ 15 (1998) 795.

\bibitem{SMMatr}N.G. Deshpande, G. Eilam,
Phys.\ Rev.\ D 26 (1982) 2463;\\
S.-P. Chia, G. Rajagopal,
Phys.\ Lett.\ B 156 (1985) 405;\\
J.M. Soares, A. Barroso,
Phys.\ Rev.\ D 39 (1989) 1973;\\
A. Barroso,
Phys.\ Rev.\ D 42 (1990) 901;\\
C.-H. Chang, X.-Q. Li, J.-X. Wang, M.-Z. Yang,
Phys.\ Lett.\ B 313 (1993) 389.

\bibitem{col} J.C. Collins,
Phys.\ Rev.\ D 58 (1998) 094002.

\bibitem{Lennart}
B.A. Kniehl, L. Zwirner, Nucl. Phys. B 621 (2002) 337.

\bibitem{FAalt}J. K\"ublbeck, M. B\"ohm, A. Denner,
Comput.\ Phys.\ Commun.\ 60 (1990) 165;\\
T. Hahn,
Comput.\ Phys.\ Commun.\ 140 (2001) 418.

\bibitem{FAFCSUSY}T. Hahn, C. Schappacher,
Comput.\ Phys.\ Commun.\ 143 (2002) 54.

\bibitem{FCLT}T. Hahn, M. P\~{e}rez-Victoria,
Comput.\ Phys.\ Commun.\ 118 (1999) 153.


\bibitem{FeynCalc}R. Mertig, M. B\"ohm, A. Denner,
Comput.\ Phys.\ Commun.\ 64 (1991) 345.

\bibitem{LoopTools}
T.~Hahn,
Acta Phys.\ Polon.\ B 30 (1999) 3469;\\
T.~Hahn,
Nucl.\ Phys.\ B (Proc.\ Suppl.) 89 (2000) 231;\\
T.~Hahn,
Nucl.\ Phys.\ B (Proc.\ Suppl.) 157 (2006) 236.

\bibitem{Cuba}T. Hahn,
Comput.\ Phys.\ Commun.\ 168 (2005) 78.

\bibitem{Hollik}A.M. Curiel, M.J. Herrero, W. Hollik, F. Merz,
S. Pe\~{n}aranda,
Phys.\ Rev.\ D 69 (2004) 075009;\\
S. Heinemeyer, W. Hollik, F. Merz, S. Pe\~{n}aranda,
Eur.\ Phys.\ J. C 37 (2004) 481.

\bibitem{AnsatzSher}
T.P. Cheng, M. Sher,
Phys.\ Rev.\ D 35 (1987) 3484.

\bibitem{Hikasa}
K. Hikasa, M. Kobayashi,
Phys.\ Rev.\ D 36 (1987) 724;\\
P. Brax, C.A. Savoy, 
Nucl.\ Phys.\ B 447 (1995) 227.

\bibitem{ciuchini}
M. Ciuchini, E. Franco, A. Masiero, L. Silvestrini, 
Phys.\ Rev.\ D 67 (2003) 075016;\\
M. Ciuchini, E. Franco, A. Masiero, L. Silvestrini, 
Phys.\ Rev.\ D 68 (2003) 079901 (Erratum).

\bibitem{PDG}Particle Data Group, W.-M. Yao et al.,
J. Phys.\ G 33 (2006) 1.

\bibitem{Denner}A. Denner,
Fortschr.\ Phys.\ 41 (1993) 307.

\bibitem{Cteq}J. Pumplin, D.R. Stump, J. Huston, H.L. Lai, P. Nadolsky,
W.-K. Tung,
JHEP 07 (2002) 012.

\bibitem{SuSpect}A. Djouadi, J.-L. Kneur, G. Moultaka,
Comput.\ Phys.\ Commun.\ 176 (2007) 426.

\bibitem{FeynHiggs}S. Heinemeyer, W. Hollik, G. Weiglein,
Comput.\ Phys.\ Commun.\ 124 (2000) 76.

\end{thebibliography}
\end{document}